\documentclass[sigconf]{acmart} 

\usepackage{xspace}
\newcommand{\system}{CardioAI\xspace}
\usepackage{xcolor} 
\usepackage{multirow}
\usepackage{graphicx}
\usepackage{booktabs}
\usepackage{array}

\newcommand{\added}[1]{\textcolor{black}{#1}} 

\newcommand\pquote[2]{{``\textit{#2}''(\textbf{#1})}}
\AtBeginDocument{%
  }

\copyrightyear{2025}
\acmYear{2025}
\setcopyright{cc}
\setcctype{by}
\acmConference[CHI '25]{CHI Conference on Human Factors in Computing
Systems}{April 26-May 1, 2025}{Yokohama, Japan}
\acmBooktitle{CHI Conference on Human Factors in Computing Systems (CHI
'25), April 26-May 1, 2025, Yokohama,
Japan}\acmDOI{10.1145/3706598.3714272}
\acmISBN{979-8-4007-1394-1/25/04}

\begin{document}

\title{CardioAI: A Multimodal AI-based System to Support Symptom Monitoring and Risk Detection of Cancer Treatment-Induced Cardiotoxicity}

\renewcommand{\shortauthors}{Wu et al.}
\renewcommand{\shorttitle}{CardioAI}


\author{Siyi Wu}
\orcid{0000-0003-2351-0049}
\authornote{Both authors contributed equally to the paper}
\affiliation{
    \institution{Northeastern University}
    \city{Boston}
    \country{USA}}

\author{Weidan Cao}
\orcid{0000-0001-5417-2121}
\authornotemark[1]
\affiliation{
    \institution{The Ohio State University Wexner Medical Center}
    \city{Columbus}
    \country{USA}}




\author{Shihan Fu}
\orcid{0009-0005-3019-6600}
\affiliation{
    \institution{Northeastern University}
    \city{Boston}
    \country{USA}}
\author{Bingsheng Yao}
\orcid{0009-0004-8329-4610}
\affiliation{
    \institution{Northeastern University}
    \city{Boston}
    \country{USA}}
\author{Ziqi Yang}
\orcid{0009-0008-8064-70002}
\affiliation{
    \institution{Northeastern University}
    \city{Boston}
    \country{USA}}

\author{Changchang Yin}
\orcid{0000-0002-6540-6365}
\affiliation{
    \institution{The Ohio State University}
    \city{Columbus}
    \country{USA}}

\author{Varun Mishra}
\orcid{0000-0003-3891-5460}
\affiliation{
    \institution{Northeastern University}
    \city{Boston}
    \country{USA}}

\author{Daniel Addison}
\orcid{0000-0002-9113-8333}
\affiliation{
    \institution{The Ohio State University Wexner Medical Center}
    \city{Columbus}
    \country{USA}}

\author{Ping Zhang}
\orcid{0000-0002-4601-0779}
\authornotemark[2]
\affiliation{
    \institution{The Ohio State University}
    \city{Columbus}
    \country{USA}}

\author{Dakuo Wang}
\orcid{0000-0001-9371-9441}
\authornote{Corresponding author.}
\affiliation{
    \institution{Northeastern University}
    \city{Boston}
    \country{USA}}

\renewcommand{\shortauthors}{Wu et al.}
\renewcommand{\shorttitle}{AI-based System to Support Symptom Monitoring and Risk Detection of Cancer Treatment-Induced Cardiotoxicity}

\begin{abstract}
Despite recent advances in cancer treatments that prolong patients' lives, treatment-induced cardiotoxicity \added{(i.e., the various heart damages caused by cancer treatments)} emerges as one major side effect. The clinical decision-making process of cardiotoxicity is challenging, as early symptoms may happen in non-clinical settings and are too subtle to be noticed until life-threatening events occur at a later stage; clinicians already have a high workload focusing on the cancer treatment, no additional effort to spare on the cardiotoxicity side effect.
Our project starts with a participatory design study with 11 clinicians to understand their decision-making practices and their feedback on an initial design of an AI-based decision-support system. 
Based on their feedback, we then propose a multimodal AI system, CardioAI, that can integrate wearables data and voice assistant data to model a patient's cardiotoxicity risk
to support clinicians' decision-making. 
We conclude our paper with a small-scale heuristic evaluation with four experts and the discussion of future design considerations. 

\end{abstract}


\begin{teaserfigure}
    \centering
    \setlength{\abovecaptionskip}{2pt}  
    \includegraphics[width=.85\linewidth]{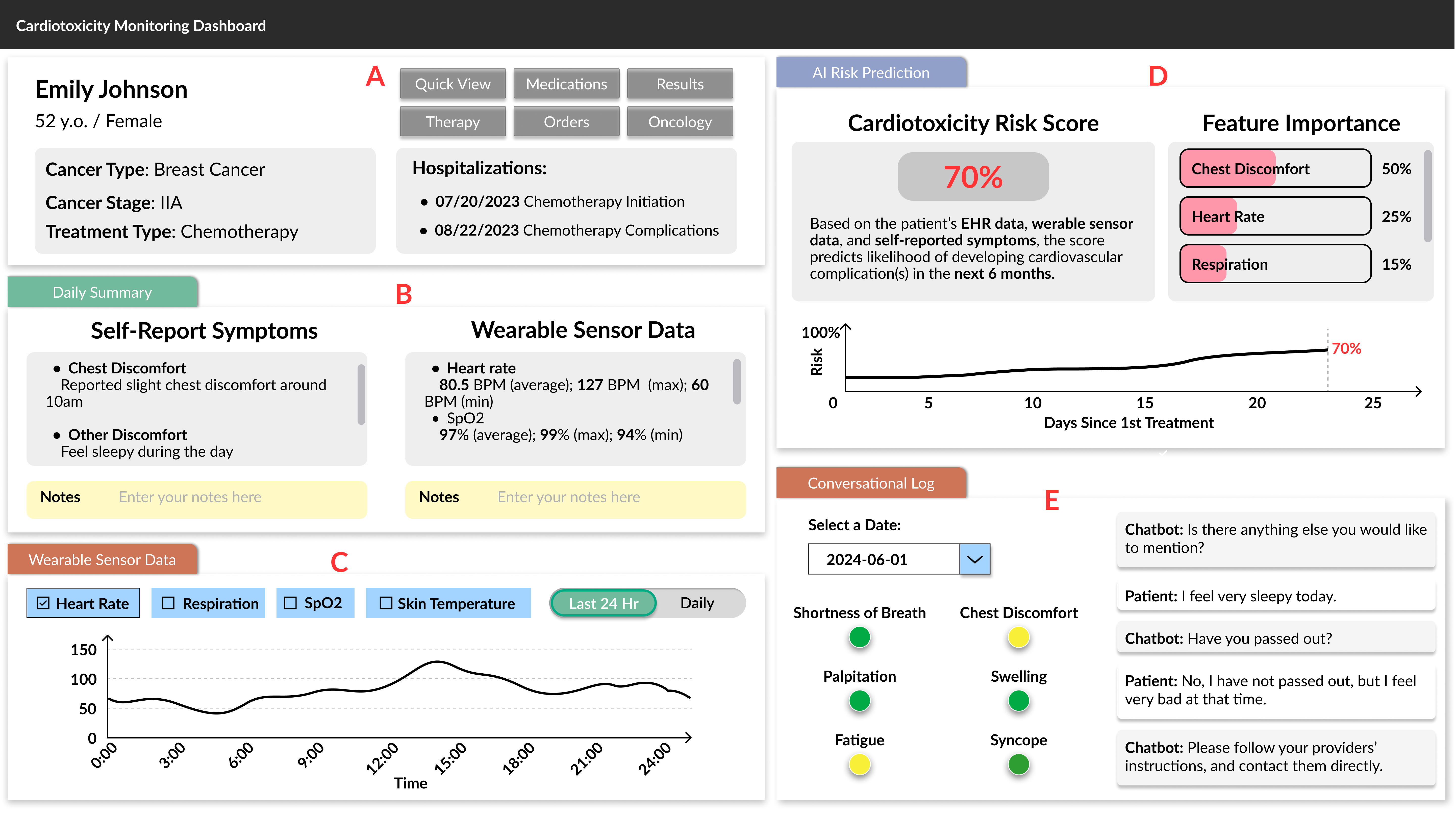}
    \caption{ \system: a multimodal AI system to support clinicians for remote monitoring and risk detection of cancer patients' cardiotoxicity risk.
    The UI has five modules: (A) Patient Information; (B) AI-generated Daily Summary; (C) Wearable Sensor Data; (D) AI-generated and Explainable Risk Score; (E) Conversation Log.
    }
    \Description{This figure depicts the main interface of the cardiotoxicity monitoring dashboard, designed to support clinicians by integrating patient data with AI predictions. The top left section, labeled A, displays the patient's details, including name (Emily Johnson), age (52 y.o.), gender (Female), cancer type (Breast Cancer), cancer stage (IIA), and treatment type (Chemotherapy). Below this, the B section presents Self-Reported Symptoms, such as chest discomfort, as well as Wearable Sensor Data, showing metrics like heart rate (80.5 BPM average) and SpO2 levels (97\% average). The Notes fields allow for additional information to be entered. In the C section, a graph shows trends in wearable sensor data, such as heart rate and respiration, over the past 24 hours. The D section to the right presents the Cardiotoxicity Risk Score, currently at 70\%, and a Feature Importance breakdown, where chest discomfort contributes 50\%, heart rate contributes 25\%, and respiration contributes 15\% to the overall risk prediction. The E section below includes a conversational log, where the chatbot asks the patient questions, and the patient responds about their current symptoms, such as feeling sleepy or reporting discomfort.}
    \label{fig:teaser}
\end{teaserfigure}

\maketitle

\section{Introduction}

Cancer treatments, such as chemotherapy, have seen remarkable advancements over the past few decades, leading to significantly improved survival rates across various types of cancer (e.g., breast cancer and sarcoma)~\cite{hu2024sequential, van2023blinatumomab, wang2013targeting, raymond2011sunitinib, schmid2018atezolizumab, borghaei2015nivolumab, hodi2010improved}. For example, thanks to advancements in chemotherapy regimens, 5-year survival rates of breast cancer patients have improved from 82.7\% to 91\%~\cite{wang2020breast}. However, advanced cancer treatments still have adverse effects (e.g., liver and kidney damage), with \textbf{treatment-induced cardiotoxicity} standing out as one of the most severe issues~\cite{truong2014chemotherapy, shaikh2012chemotherapy, florescu2013chemotherapy}. Treatment-induced cardiotoxicity refers to the various heart damage caused by cancer treatments, which can lead to minor or severe consequences such as heart failure, arrhythmia, and other cardiovascular diseases~\cite{alvi2019cardiovascular, baptiste2019high}. To put this into the breast cancer treatment context, cardiotoxicity has emerged as the leading cause of morbidity and mortality in long-term follow-up~\cite{henry2018cardiotoxicity}. 

\added{Recent studies have investigated early decision-making of treatment-induced cardiotoxicity~\cite{curigliano2016cardiotoxicity, conway2015prevention}. However, this is challenging due to the following four reasons. First, the initial symptoms of cardiotoxicity are often subtle in the early stages, 
and often, by the time clinicians are able to identify by obvious symptoms, cardiotoxicity has already caused irreversible damage~\cite{kalam2013role, lima2022cardiotoxicity}
~\textbf{(Challenge 1)}\label{challenge1}. 
Second, symptoms can manifest outside clinical settings, where patient-clinician interactions are significantly reduced, causing symptoms to go unnoticed or be diagnosed late~\cite{henry2008symptoms}~\textbf{(Challenge 2)}\label{challenge2}. 
Third, the current healthcare system heavily relies on patient self-reporting for symptom tracking outside the hospital, but patients may falsely report and have poor compliance due to lack of health literacy and forgetfulness~\cite{cardinale2000left, cardinale2006prevention, pai2000cardiotoxicity}~\textbf{(Challenge 3)}\label{challenge3}. 
Lastly, clinicians face high workload and pressure, which limits their capacity to process massive amounts of clinical information in a short period of time necessary for early cardiotoxicity detection~\cite{ewer2010cardiotoxicity, curigliano2016cardiotoxicity}~\textbf{(Challenge 4)}\label{challenge4}. While these challenges have been previously identified, their specific impact on clinicians’ decision-making processes remains underexplored. 
Without addressing these challenges, the identification of cancer treatment-induced cardiotoxicity is very difficult and often delayed, which could lead to belated disease intervention, and significantly impact overall patient outcome~\cite{wu2024clinical}.}

To mitigate these challenges, we aim to develop an AI-based system to support clinicians' early decision-making of cancer treatment-induced cardiotoxicity. 
Two groups of literature are particularly inspiring: \textbf{Remote Patient Monitoring (RPM)}~\cite{calambur2024case,nguyen2023toward,zhang2023experiences} and \textbf{AI-based Clinical Decision Support Systems (AI-CDSS)}~\cite{wang2023human}. \textbf{RPM} uses technologies (e.g., wearable devices or home sensors) to continuously collect patient health data (e.g., blood pressure) outside of clinical settings~\cite{catalyst2018telehealth,us2008national}, which could be a promising solution to cope with the aforementioned challenges of monitoring non-clinical subtle early symptom data~\added{(\textbf{Response to Challenge 1 \& 2)}}.
Furthermore, recent HCI research has shown the potential of using LLM-based voice assistants (LLM-VA) as verbal communication interfaces to support patients to self-report symptoms~\cite{mahmood2023llm, zhou2024performance, yang2024wish, li2024scoping}, which can be a powerful complement to traditional \added{self-reporting approaches} for our user case \added{~(\textbf{Response to Challenge 3})}. With all the multimodal data continuously collected via RPM and LLM-VA, it may pose an additional cognitive demand on the already high-workload clinician decision makers; that is where
\textbf{AI-CDSS} may help by automatically organizing and analyzing the raw data, and presenting only a high-level risk prediction to the clinicians~\cite{chiang2023two} \added{~(\textbf{Response to Challenge 4})}. 

\added{One key issue of designing a successful AI-CDSS is that the AI needs to be explainable and trustworthy to clinicians, and it should integrate into their existing workflow.} 
Some existing HCI literature~\cite{plana2014scherrer,sepsis,albahri2023systematic} exemplified the best practices such as following the human-centered AI (HCAI) principles in design~\cite{amershi2019guidelines} and engaging stakeholders throughout the design process~\cite{wang2023human,sepsis}, which inspired our methodology. In this paper, we propose a first-of-its-kind multimodal AI-based system that combines the strengths of RPM (continuous monitoring multimodal symptom data via wearables and a LLM-VA \added{to address \textbf{Challenge 1, 2, and 3}}) and explainable AI-CDSS (\added{designed to mitigate \textbf{Challenge 4} by offering} a model-based predictive risk score and explanation generation). The design process also follows HCAI design principles~\cite{amershi2019guidelines}. In particular, our work aims to answer two research questions:

\begin{itemize}
    \item RQ1: How do clinicians’ experiences provide deeper insights into the challenges of early decision-making for cancer treatment-induced cardiotoxicity, and what technological design needs do they have? 
    \item RQ2: How will clinicians perceive and interact with our initial design of a multimodal AI-based system in early decision-making of cancer treatment-induced cardiotoxicity?
\end{itemize}

\added{From RQ1, we aim to gather clinicians' challenges and technological needs in today's decision-making workflow for cancer treatment-induced cardiotoxicity. 
To answer RQ1, we conducted participatory design (PD) sessions with 11 clinicians, fostering in-depth discussions and actively involving them in the co-design process. During each session (Figure~\ref{fig:first-UI}), we presented an initial design sketch  to gather feedback and suggestions. 
Through this collaborative and iterative process, we developed our pilot system, CardioAI, which is detailed in Section~\ref{sec:4}. CardioAI offers two main functionalities: (1) ~\textbf{Symptom Monitoring}: vital signs collected from wearables (Module C in Figure ~\ref{fig:teaser}) and qualitative symptoms self-reported via an LLM-VA (Module E); (2) ~\textbf{Risk Prediction and Summarization}: AI-generated predictive risk scores for cardiotoxicity paired with explanations as feature importance (Module D) and summarizations of the collected data (Module B).}

\added{Lastly, to address RQ2, we conducted a heuristic evaluation with four clinicians, including two cardiologists and two oncologists, to gather feedback on the usability and functionality of our pilot system. 
Data were collected through cognitive walkthroughs and think-aloud protocols during clinicians' interactions with the system, along with post-study interviews and usability surveys.  Clinicians highlighted the system's ability to provide additional cardiotoxicity-specific patient data for monitoring and diagnosis, reduce cognitive workload, and support proactive decision-making within their existing workflows.}

In summary, our contributions are as follows:
\begin{itemize}
    \item Following the human-centered design principles and practices, we co-designed with \added{clinicians}  a multimodal AI-based \added{pilot} system that integrates continuous \added{symptom} monitoring and \added{explainable} AI model to support early decision-making of cancer treatment-induced cardiotoxicity. 
    \item Through a follow-up user evaluation of our prototype, we identified key design considerations for future implementations of multimodal AI-based systems in clinical settings.
\end{itemize}

\section{Related Work}

\subsection{Challenges in Cancer Treatment-Induced Cardiotoxicity}~\label{challenges}

Despite the survival rate of cancer patients has improved significantly due to advances in cancer treatment development, new concerns and risks emerge alongside them, such that treatment-induced cardiotoxicity is a particular complication that has aroused more and more attention~\cite{sant2009eurocare, sant2009eurocare, oeffinger2006chronic}.
Treatment-induced cardiotoxicity represents damage to the cardiovascular system, such as heart failure~\cite{herrmann2020adverse, schlitt2014cardiotoxicity}, caused by cancer treatments, which poses serious risks to patients' well-being during and after cancer treatment~\cite{kalam2013role, gharib2002chemotherapy}. 
Cardiotoxicity induced by anthracyclines, in particular, can lead to a two-year mortality rate of up to 60\%~\cite{swain2003congestive, saini1987reversibility, von1979risk}
The development of cardiotoxicity can also lead to interruption of cancer treatment if the vital organs of the patients cannot endure the side effects of subsequent cancer treatment~\cite{lima2022cardiotoxicity}.
Thus, monitoring cardiotoxicity at an early stage is crucial, as patients can benefit from early diagnosis and intervention for cardiotoxicity to reduce the risk of dangerous negative effects~\cite{cardinale2000left, cardinale2006prevention, pai2000cardiotoxicity}.

However, detecting early signals of treatment-induced cardiotoxicity for timely intervention is a significant challenge in the healthcare field~\cite{cardinale2010anthracycline, kalam2013role}. One major issue is that initial cardiac symptoms are often very mild, leading to delayed self-reporting from patients~\cite{wittayanukorn2017cardiotoxicity, mascolo2021immune, kamphuis2019cancer}~\added{(Challenge 1)}. When more pronounced symptoms (e.g., edema) appear, the condition may have already progressed to an irreversible state~\cite{kalam2013role, lima2022cardiotoxicity}. This delay hinders clinicians' ability to diagnose cardiotoxicity early during and after cancer treatment~\cite{kalam2013role}. Additionally, treatment-induced cardiotoxicity can manifest at varying times—during treatment, shortly after, or even years later~\cite{pai2000cardiotoxicity}. This variability makes consistent monitoring challenging, especially when symptoms arise outside the hospital setting~\cite{sheppard2013cardiotoxicity}~ \added{(Challenge 2)}.
The reliance on patient self-reporting is another critical challenge. Patients may underreport or misreport symptoms due to forgetfulness, misunderstanding, or downplaying their discomfort~\cite{cardinale2000left, cardinale2006prevention, pai2000cardiotoxicity}. This underreporting can lead to missed early signs of cardiotoxicity, delaying necessary interventions~\added{(Challenge 3)}.

Furthermore, clinicians are frequently tasked with the demanding responsibility of overseeing cancer treatments~\cite{ewer2010cardiotoxicity, curigliano2016cardiotoxicity}, leaving them with little capacity to rapidly gather and analyze extensive clinical data, which is an essential step for the early detection of cardiotoxicity~\cite{wu2024clinical}~\added{(Challenge 4)}. 
There is also a lack of standardized guidelines for clinicians to detect and intervene in treatment-induced cardiotoxicity effectively~\cite{talty2024home, moustafa2023critical}, which exacerbates the difficulty in early detection and consistent management across healthcare providers. Due to these compounded challenges, there is a high demand for novel approaches that offer effective and efficient monitoring and diagnosis of treatment-induced cardiotoxicity~\cite{gharib2002chemotherapy}.

\subsection{Remote patient monitoring for Cancer Care}~\label{RPM}
Remote patient monitoring (RPM) involves leveraging digital technologies to collect health data from patients outside clinical settings and transmit this information to healthcare providers for evaluation ~\cite{catalyst2018telehealth,us2008national}. 
In cancer care, RPM has greatly improved clinicians' ability to monitor and manage patient health, providing a cost-effective solution~\cite{vegesna2017remote, kofoed2012benefits}. 
Such improvement is driven by the integration of telehealth systems~\cite{8857717}.
Telehealth systems, including the use of technology-driven platforms to provide different types of health-related information and services~\cite{mechanic2017telehealth}, have facilitated cancer patients' access to healthcare consultations and personalized coaching, boosting patient engagement and adherence to treatment protocols~\cite{larson2018effect, cox2017cancer, chen2018effect}. 
Patients can self-administer their telehealth systems, and data can be recorded and transmitted to clinicians remotely using a mobile phone~\cite{doyle2020state, grewal2020telehealth}.
For instance, wearable telehealth systems~\cite{dias2018wearable, shaik2022fedstack} are increasingly adopted in clinical settings to provide continuous real-time monitoring of patients' physiological data, offering more accuracy than self-reported symptoms~\cite{hardcastle2020fitbit, gresham2018wearable, baig2017systematic}. 
For example, fitness trackers and smartwatches can monitor vital signs that are critical for the early detection of cardiotoxicity, such as heart rate, physical activity levels, sleep patterns, oxygen saturation, and respiratory rates, which can offer a thorough understanding of patients' health status~\cite{hardcastle2020fitbit, tadas2023using}.

While these telehealth systems have shown effectiveness in monitoring specific aspects of cancer patient health conditions remotely, they only focus on collecting physiological data that can be captured by sensors. 
In contrast, nurse practitioners make regular phone calls after cancer treatment to collect patients' self-reported feelings, and nurses can ask follow-up questions to correspond to patients' feedback to better understand their well-being. 
Natural language communication between nurses and patients comprises rich information regarding the patient's physiological wellness that cannot be captured by existing sensor-based telehealth systems~\cite{raschi2010anticancer}.
Thus, increasing research attention has been focused on the development of telehealth systems with verbal communication interfaces, such as conversational agents, that support gathering self-reported patient health data, such as daily well-being status for cancer care~\cite{larson2018effect, kocaballi2019personalization}. 
For instance, the Nurse AMIE project uses smart speakers to deliver mental and informational care interventions to women with metastatic breast cancer~\cite{qiu2021NurseAMIE}; \citet{gregory2023exploring} designed a mobile health prototype to track the cardiac symptoms of cancer patients using questionnaires.
Researchers have also developed conversational agents for a number of clinical inquiries of cancer patients, including clinical diagnosis, patient education, and symptom monitoring~\cite{zhou2024performance, li2024scoping, yang2024talk2care, xu2021chatbot, kim2024mindfuldiary}.

A thorough understanding of cancer patients' well-being outside the clinical setting is particularly crucial for the monitoring and diagnosis of treatment-induced cardiotoxicity because of its variability in onset and high-stake nature. 
In particular, physiological data collected via wearable devices can provide real-time insight into subtle changes in cardiac function that might go unnoticed by the patient, whereas telehealth systems with verbal communication interfaces can capture rich health information in natural language self-reflections by patients. 
In this work, we propose an AI-based system to collect multimodal data with the help of wearable devices and voice assistants to address the needs for early monitoring and identification of treatment-induced cardiotoxicity~\cite{shelburne2014cancer, sheppard2013cardiotoxicity}.

\subsection{Design of AI-Supported Clinical Decision Support Systems}~\label{Decision-making}

Recent advances in AI predictive modeling shed light on a broad avenue for supporting clinicians' clinical decision-making with accessible and efficient AI assistance~\cite{sepsis}.  
Such technical advancement, in turn, led to a growing research interest from HCI experts and healthcare practitioners in the design of AI-CDSS~\cite{chiang2023two, changArtificialIntelligenceApproach2022, lalArtificialIntelligenceRisk2023, ahmed2024advancements, kwanMultimodalityAdvancedCardiovascular2022, yagiArtificialIntelligenceenabledPrediction2024}.
AI-CDSS systems can greatly improve the efficiency of clinicians' clinical workflow, where clinicians have to process a large amount of complicated electronic health records (EHRs) data to make informed clinical decisions~\cite{ewer2023cardiac, steinhubl2013can, steinhubl2015emerging, free2013effectiveness}. 
However, traditional AI-CDSS approaches rely on clinical results for risk prediction or serve only general-purpose communication while facing reliability, accuracy, or privacy concerns in AI applications in healthcare care~\cite{changArtificialIntelligenceApproach2022, lalArtificialIntelligenceRisk2023, ahmed2024advancements}.

Specifically, in the context of treatment-induced cardiotoxicity, AI has the potential to facilitate clinicians in making more accurate diagnoses and, ultimately, in providing appropriate therapeutic interventions~\cite{moazemi2023artificial, yin2024sepsislab}.
For example, \citet{changArtificialIntelligenceApproach2022} trained AI with clinical, chemotherapy, and echocardiographic parameters to predict cancer treatment-induced cardiac dysfunction (CTRCD) and heart failure, reaching higher accuracy than traditional prediction models; \citet{yagiArtificialIntelligenceenabledPrediction2024}'s model robustly stratified CTRCD risk from baseline electrocardiograms (ECG). 
However, current AI-CDSS solutions for cardiotoxicity detection face several limitations. 
One key limitation is that these models often rely on EHR data collected during infrequent clinical visits, resulting in incomplete and limited datasets that may miss subtle, early indicators of cardiotoxicity~\cite{plana2014scherrer}, which hinders the development of accurate predictive models.
Furthermore, existing AI systems lack integration with the principles of HCAI, which hampers their effectiveness in real-world healthcare settings~\cite{sepsis}. 
For example, they often do not fit seamlessly into existing clinical workflows and can lead to information overload if too much information is presented to clinicians at the same time, leading to inefficiencies and frustration for healthcare providers. 
To bridge the gap, our work aims to explore how to efficiently and effectively monitor rich multimodal patient data with RPM technologies, and leverage these data to enable AI-CDSS to support clinicians' decision-making for cardiotoxicity early detection.

\section{Participatory Design}
\label{Sec:3}

\added{Although prior research has identified four key challenges in managing cancer treatment-induced cardiotoxicity and explored the potential of emerging technologies, clinicians’ perspectives on specific design requirements for integrating these technologies into clinical workflows remain underexplored. To bridge this gap, we conducted a participatory design (PD) study with clinical experts specializing in cancer treatment-induced cardiotoxicity, holding sessions individually. The study aims to (1) map how the opportunities offered by emerging technologies can address these challenges and (2) understand the specific needs of clinician technology design to ensure effective integration into their decision-making workflows. By focusing on clinicians’ insights, we sought to align our system design with their practical needs and requirements.}
\begin{table*}[!h]
    \centering
    \resizebox{\textwidth}{!}{
    \begin{tabular}{c|c|c|c|c}
    \toprule
      \textbf{P\#} & \textbf{Gender} & \textbf{Department} & \textbf{Job Title} & \textbf{Year of Practice}\\
    \midrule
      P1  &  Female & Medical Oncology  & Breast Medical Oncologist & 2 years\\
      P2  &  Male & Radiation Oncology & Gastrointestinal Radiation Oncologist & 6 years\\
      P3  &  Female & Bone and Marrow Transplant and Cellular Therapy & Hematologist& 1.5 years\\
      P4  &  Female &  Internal Medicine    &  Oncologist & 13 years\\
      P5  &  Female & Sarcoma Center    & Sarcoma Medical Oncologist & 2 years\\
      P6  &  Female & Thoracic and Geriatric Oncology   & Thoracic and Geriatric Oncologist & 7 years\\
      P7  &  Male & Internal Medicine   & Cardiologist &  8 years\\
      P8 & Female & Internal Medicine & Hematologist & 7 years\\
      P9 & Male & Hematology& Hematologist& 4 years \\
      P10 & Male & Medical Oncology & Thoracic Medical Oncologist & 7 years\\
      P11 & Male & Internal Medicine& Cardiologist& 7 years\\
    \bottomrule
    \end{tabular}
    }
    \caption{Demographics of Participants in Our PD Study}
    \label{tab:interviewees}
\end{table*}

\subsection{Study Participants}
\label{sec3.1}

\added{The PD study aimed to explore how technologies could be tailored to clinicians’ needs, specifically to support their decision-making processes in managing treatment-induced cardiotoxicity. To achieve this, we recruited 11 clinicians with significant experience in this area by convenience sampling~\cite{sedgwick2013convenience}. Recruitment involved leveraging professional networks, referrals from colleagues, and participant recommendations, ensuring a diverse pool of experts. All participants were affiliated with a North American hospital, representing key specialties -- oncology and cardiology -- that are central to decision-making for cancer treatment-induced cardiotoxicity. 
Table~\ref{tab:interviewees} presents the demographics of the participants with a randomly assigned number denoted as $P\#$. 
Participants came from various departments and had a wide range of clinical experience, spanning 1.5 to 13 years. }

\subsection{Study Procedure}
\added{The PD sessions were structured to actively involve clinicians in identifying challenges, exploring technological opportunities, and co-designing solutions to integrate emerging technologies into their workflows. Each individual session was divided into three phases: (1) entry questions, (2) participatory design with an initial User Interface (UI) draft, and (3) exit questions.}

\begin{figure*}[!tp]
    \centering
    \includegraphics[width=1\textwidth]{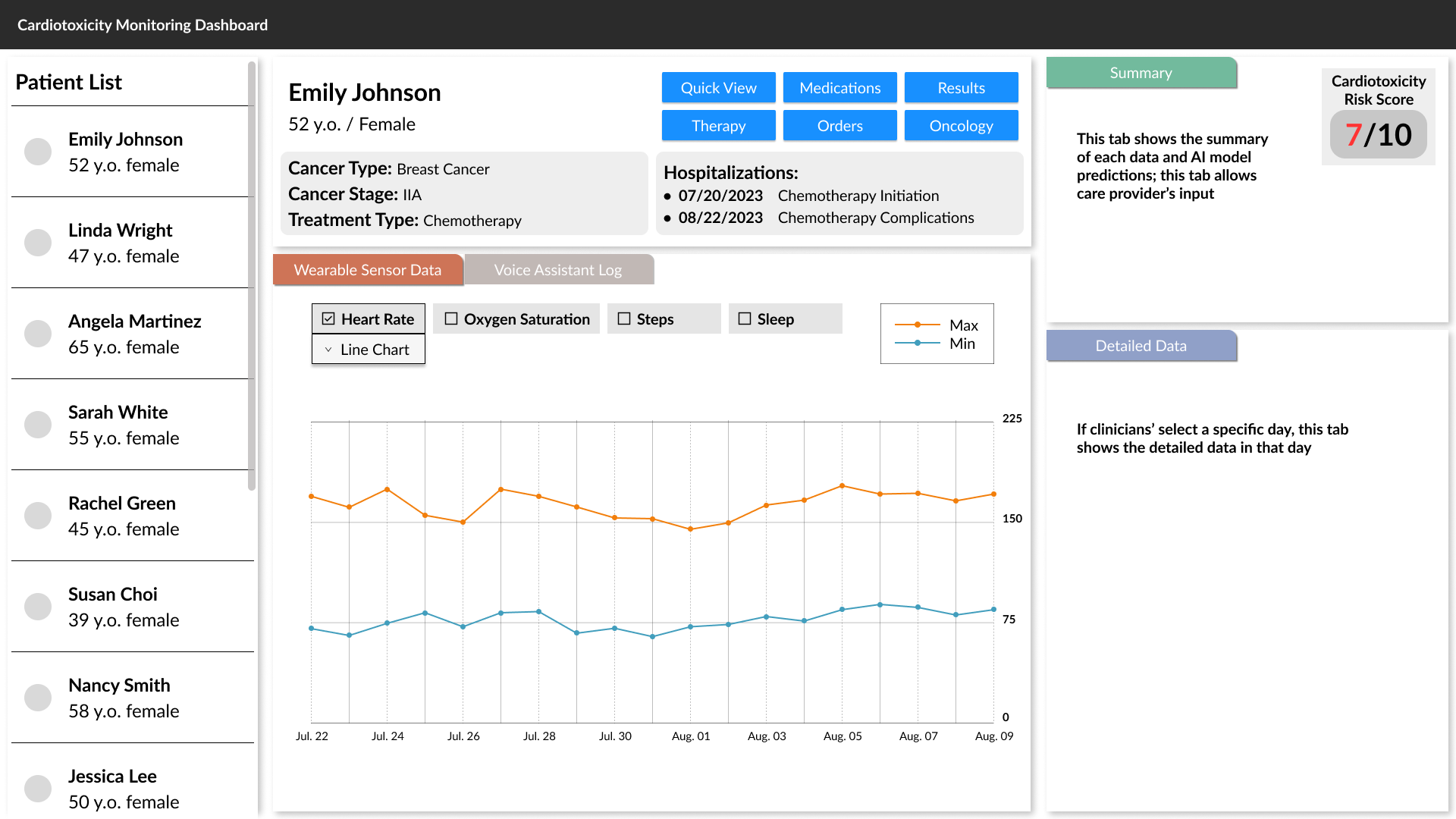}
    \caption{Our participatory design session. A participant is suggesting design revisions on the initial UI.}
    \Description{A screenshot showing the initial user interface prototype for a clinician-facing cardiotoxicity monitoring dashboard, presented during a participatory design session. On the left side of the screen, a Patient List displays the names and details of patients being monitored, including their age and gender. The selected patient, Emily Johnson, has her profile displayed on the right side. This includes her cancer type (Breast Cancer), stage (IIA), and treatment type (Chemotherapy), along with hospitalization dates. The dashboard provides options for viewing Wearable Sensor Data, a Voice Assistant Log, and patient-specific data visualizations. In the center, a graph displays trends in heart rate and oxygen saturation over time, with maximum and minimum values tracked for each metric. Additional buttons allow clinicians to select and view other types of data, such as steps and sleep. On the right side of the screen, a Cardiotoxicity Risk Score of 7/10 is shown, along with options to view Summary and Detailed Data tabs. The summary presents an overall view of the patient’s data and AI model predictions, while the detailed data tab allows clinicians to drill down into specific data for a selected day.}
    \label{fig:first-UI}
\end{figure*}

\begin{figure*}[!tp]
    \centering
    \begin{minipage}{0.32\textwidth}
        \centering
        \includegraphics[width=\textwidth]{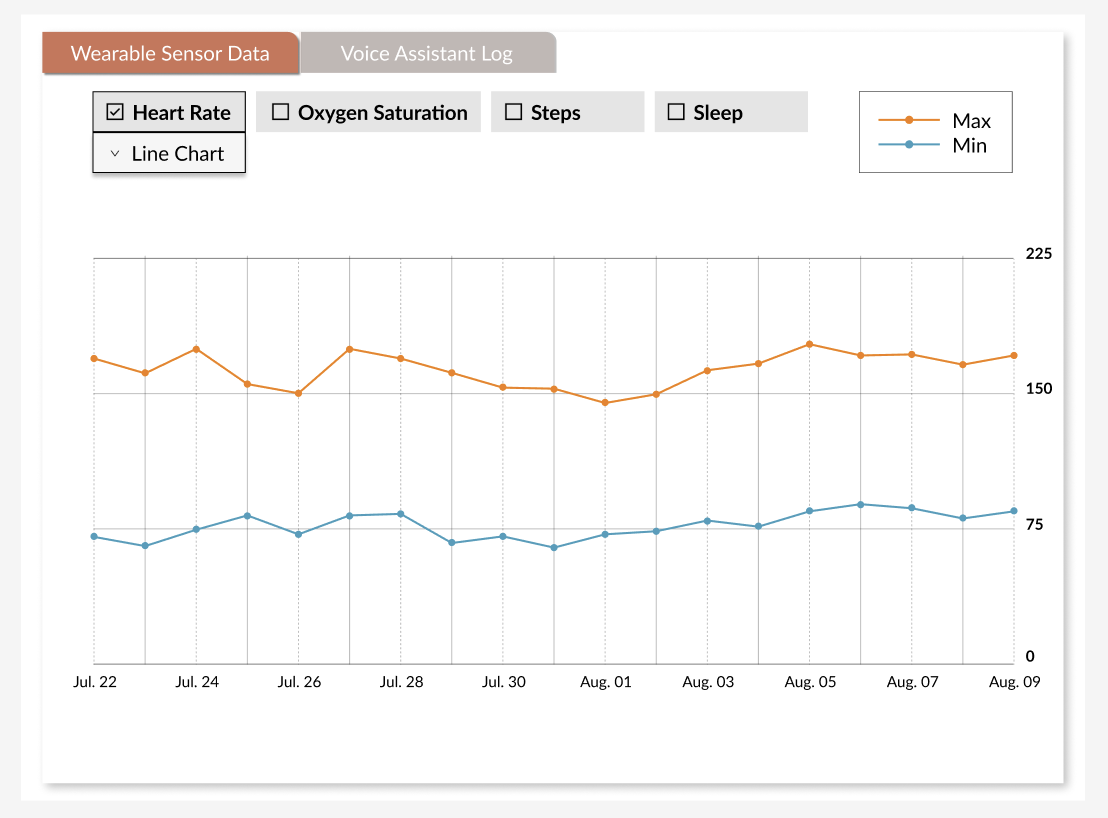}
    \end{minipage}
    \hfill
    \begin{minipage}{0.32\textwidth}
        \centering
        \includegraphics[width=\textwidth]{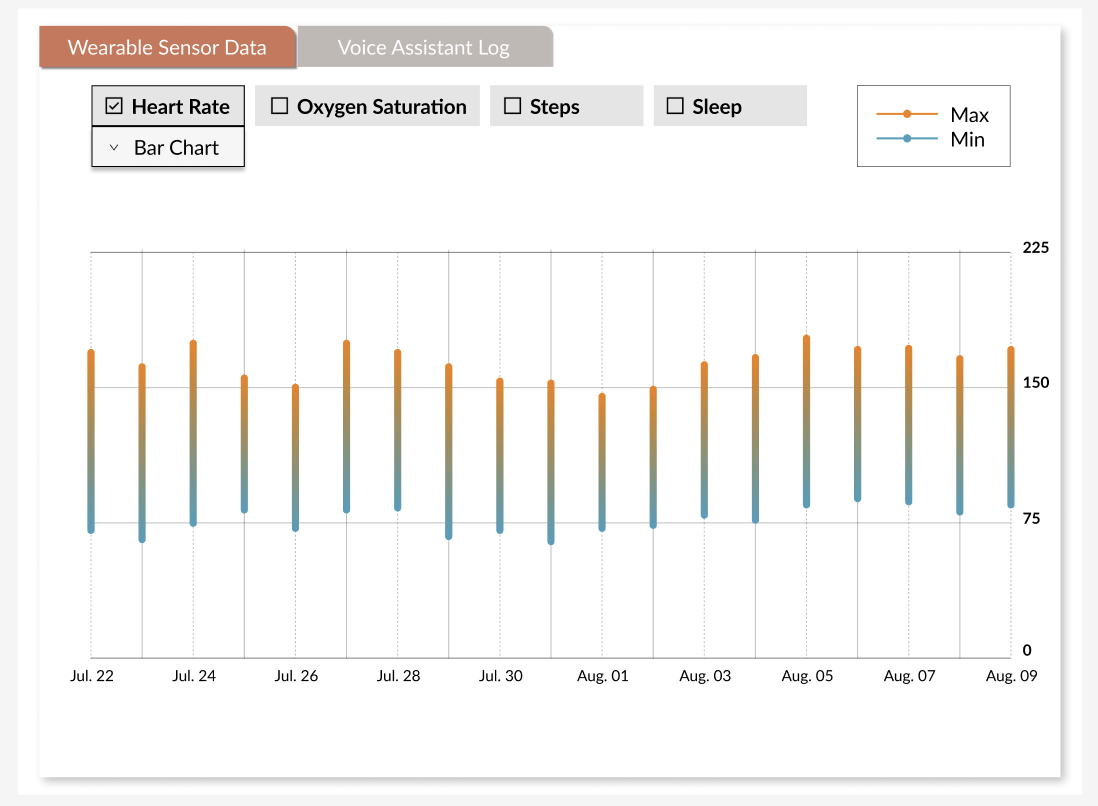}
    \end{minipage}
    \hfill
    \begin{minipage}{0.32\textwidth}
        \centering
        \includegraphics[width=\textwidth]{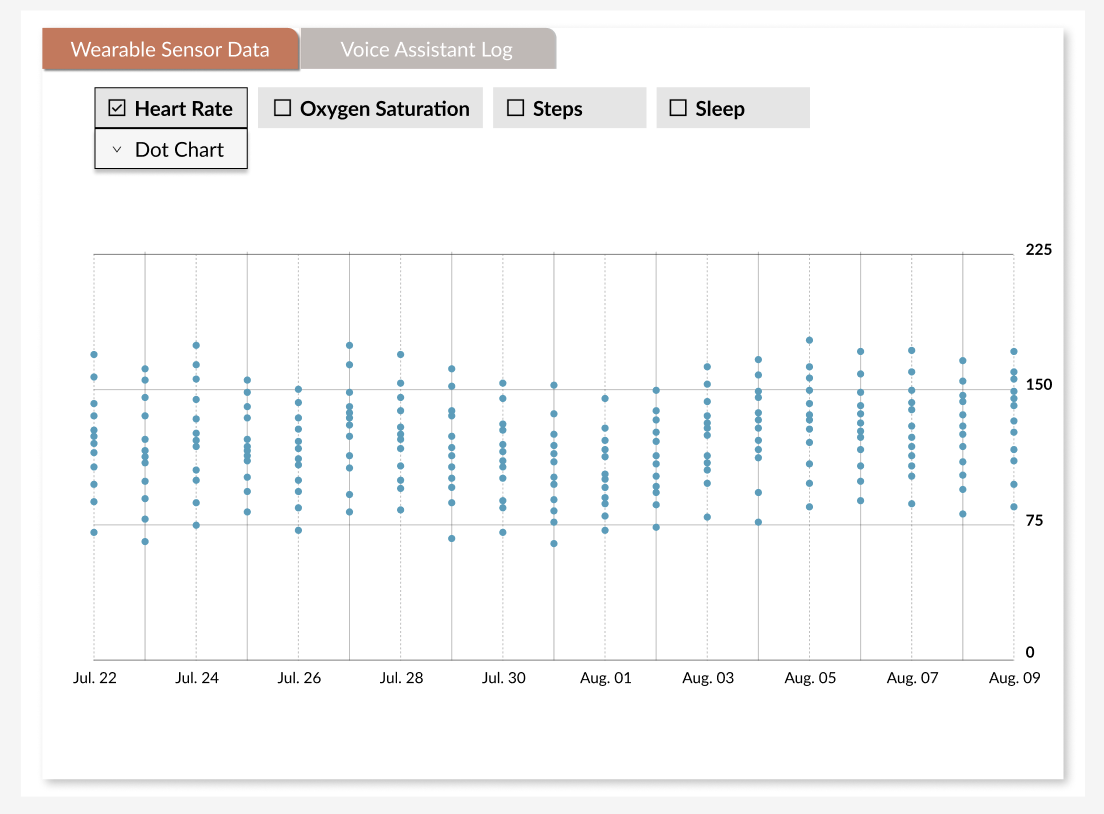}
    \end{minipage}
    \caption{Examples of low-fidelity visualization options for heart rate data used to elicit clinician feedback.}
    \Description{A set of three side-by-side low-fidelity visualization options for heart rate data designed to elicit clinician feedback during participatory design sessions. The first visualization is a line chart showing trends in maximum and minimum heart rate values over time. The second visualization is a bar chart displaying the same maximum and minimum heart rate values across different time points. The third visualization is a dot chart representing individual heart rate data points, distributed along a timeline. Each chart is labeled "Wearable Sensor Data" with selectable options for additional metrics, such as "Oxygen Saturation" and "Steps.”}
    \label{fig:three-images}
\end{figure*}

\added{We aimed to ensure all participants had a shared conceptual understanding of the technologies explored in RPM and AI-CDSS literature. To achieve this, before PD sessions, we developed an initial low-fidelity UI draft~(Figure~\ref{fig:first-UI}) for a clinician-oriented information dashboard. This prototype served as a probe to facilitate the co-design process, focusing on eliciting participants' broad feedback. Each feature was represented with placeholders or illustrative examples to demonstrate potential outputs and functionalities, helping bridge gaps in technical literacy and creating a shared foundation for discussion. 
Rather than presenting a fixed design, we offered multiple design options, varying in both content (e.g., visualizing different types of physiological data such as heart rate and oxygen saturation) and design (e.g., using line charts, bar charts, and dot plots). Examples of such design options are included in Figure~\ref{fig:three-images} as prompts for participant feedback.
The initial UI design was informed by prior literature and included conceptual representations of key features such as wearable devices, LLM-VA, and AI-driven prediction models. These conceptual elements were intended to spark discussions around potential functionalities, interactions, and design considerations for integration into clinical workflows.}

\added{At the start of each session, participants were asked to recall a recent case involving cancer treatment-induced cardiotoxicity, contextualizing the challenges in monitoring and detection within their workflows. Following this, participants were asked to describe their current use of technologies in these workflows, share their familiarity with emerging tools, and provide their perspectives on potential technologies being studied or proposed for decision-making support in this domain.}

\added{During the PD phase, participants actively contributed to the refinement of the initial UI draft. They asked clarifying questions, proposed design revisions, and suggested new features with specific examples. They also envisioned practical use cases for particular scenarios and raised concerns about potential limitations. We concluded each PD session with exit questions to gather final reflections and additional insights from participants. All interview questions can be found in Appendix ~\ref{appendix:interviewQ}. }

\added{The sessions were conducted remotely via Zoom by the first author and lasted 35 to 50 minutes. All sessions were recorded and transcribed with the participants’ consent. Two members of the research team independently analyzed the transcripts using inductive coding and thematic analysis~\cite{braun2012, braun2019}. Through iterative discussions, the codes were refined into a consensus codebook to ensure an accurate representation of the data, minimizing redundancy and merging overlapping themes. The transcripts were re-coded based on the finalized codebook (Appendix ~\ref{Appendix:codebook}). 
During the analysis, we observed that thematic saturation was naturally achieved. As we continued coding, no new themes emerged from the data after analyzing the transcripts of the 11 participants.}

\added{Our research team combined expertise in HCI, clinical practice (with specialized knowledge in cancer treatment-induced cardiotoxicity), and AI/ML, enabling a comprehensive understanding of our participants. We acknowledge that our interdisciplinary perspectives influenced the design, analysis, and interpretation of the study. By adopting a user-centered approach, we aimed to prioritize the participants’ voices while using our diverse expertise to contextualize their insights within the broader domain.}

\subsection{Findings} 
\label{sec3.2}

\subsubsection{Challenges in Clinical Decision-making for Cancer Treatment-Induced Cardiotoxicity}
\label{sec3-challenges}
Previous studies~\cite{cardinale2010anthracycline, kalam2013role, kalam2013role, wittayanukorn2017cardiotoxicity, mascolo2021immune, kamphuis2019cancer, pai2000cardiotoxicity, sheppard2013cardiotoxicity, wu2024clinical} have highlighted 
\added{four main challenges in decision-making for cancer treatment-induced cardiotoxicity as we described in Section ~\ref{challenges}. While many of the challenges our participants described align with these four, they provide more nuanced insights into them. }

\added{\textit{Symptoms Not just Subtle and Infrequent, But Sometimes Absent in the Early Stages.}}
\added{Our participants reinforced \textbf{Challenge 1\&2} that symptoms are often subtle and infrequent in the early stages and difficult to detect during routine clinical assessments or outside of clinical settings. They also highlighted a critical nuance: in some cases, symptoms may be entirely absent, even when patients are experiencing severe conditions. For example, P4 shared a case where a patient, completely asymptomatic at home, was found to have a severe arrhythmia solely because he happened to wear a heart monitor, a device not routinely provided to all patients:
}

\begin{quote} \pquote{P4}{He had been on the heart monitor for about 3 or 4 days... I got this message he had a severe form of arrhythmia... He was completely asymptomatic. He was sitting at home relaxing, and the monitor picked up this thing.} \end{quote}

\added{\textit{Logistical Barriers and Non-Specific Tools Compound Self-Reporting Limitations.}}
\added{Participants reiterated \textbf{Challenge 3}, emphasizing that self-reporting, the primary means through which clinicians learn about patients' symptoms at home, is fraught with limitations. They also pointed out factors that exacerbate this challenge, including logistical barriers and design-level limitations, which extend beyond the known issue of health literacy and patients' tendency to downplay symptoms.}
\added{One issue raised by participants is the presence of logistical barriers, which possibly hinder the effectiveness of self-reporting even when patients are aware of their symptoms.} These barriers, such as delayed communication between patients and healthcare providers, are particularly pronounced in resource-constrained settings, leading to discouragement and increased cognitive load on patients, as described by P1:  
\begin{quote} \pquote{P1}{In some areas, patients may not hear back right away from their physicians or nurses if they’re having symptoms... They may feel like they’re complaining too much or being a burden.} \end{quote}

\added{Participants also revealed} the difficulty in recognizing cardiotoxicity-specific symptoms within current self-reporting systems.
\added{Many tools currently in use are designed for general purposes rather than tailored to the nuances of cardiotoxicity. This gap leads to missed or underreported cardiac symptoms.} For example, P5 emphasized that the surveys they use are \textit{``not specific to cardiotoxicity... unless there’s anything on the exam that suggests a cardiac problem.''}

\added{\textit{Absence of Risk patterns as an Additional Burden to Clinician Workload.}}
Several participants (P1, P2, P7) expressed their frustration with the absence of reliable patterns or correlations between patient characteristics and cardiotoxicity risk, making it challenging to determine which patients require more intensive monitoring. 
As P1 noted, ``\textit{We don’t have a good grasp on what features correlate with the likelihood of someone developing cardiotoxicity.}''
Such uncertainty often results in reactive decision-making, where clinicians have to wait for symptoms to manifest before intervening rather than being able to anticipate and address risks sooner. \added{The uncertainty surrounding risk factors further complicates an already demanding workload \textbf{Challenge 4}, leaving clinicians to manage the additional cognitive and logistical burden of deciding which patients need closer monitoring.}
Such delays, compounded by late symptom reporting and diagnostic processes, can lead to belated intervention. 
P2 highlighted this issue by sharing a case in which delays in reporting and testing prolonged treatment decisions: ``\textit{The symptoms were reported late... everything was delayed, from the echo to the stress test to the EKG.}''

\subsubsection{Opportunities and Suggested Design Revisions}
\label{sec3.2.2}

\added{To explore the potential of technologies that can assist clinicians in addressing the identified challenges, we began by asking about their familiarity with wearables, LLM-VA, and AI-based predictive risk scores. We then introduce these concepts alongside an initial UI draft of a clinician-facing dashboard that integrates these modules to facilitate discussion. Participants expressed strong interest in the system's potential, highlighting key opportunities, sharing their perceptions, and providing suggestions for design improvements. We summarize the key insights below.}

\added{\textbf{Opportunity 1: Continuous Monitoring of Key Clinical Metrics Using Wearables.}}
\added{Participants highlighted the potential of wearables to provide continuous, real-time monitoring of patient's vital signs, such as heart rate and blood pressure, which could enable identifying subtle patterns and anomalies in patient health. Reflecting on their experiences, clinicians noted the value of such monitoring, with P4 commenting,}
\added{\textit{``I think we can pick up more patients with this. Or we can pick up patients earlier with this... We may be able to find more cardiac patients or cardiac adverse events related to our medications by doing this kind of monitoring.''}.
Continuous monitoring was also seen as a way to mitigate situational factors in clinical settings, such as ``white coat hypertension'', where patients exhibit elevated blood pressure due to anxiety during clinical visits. This opportunity provided by wearables could help clinicians recognize early and subtle symptoms, potentially responding to \textbf{Challenge 1}.}

\added{Building on this opportunity, clinicians further emphasized the importance of continuously monitoring a
comprehensive set of physiological parameters that are clinically important to cardiotoxicity, which leads to our \textbf{Design Revision 1: Monitoring Key Clinical Symptoms}. Many clinicians mentioned heart rate, blood pressure, oxygen saturation, and, where possible, additional indicators such as blood sugar levels. They were curious about the potential of wearable devices to track these diverse metrics, particularly
in patients with complex health profiles. As P1 explained,}
\added{\begin{quote}
\pquote{P1}{What other things could you track with the wearable? Blood sugar would be another one if that would be possible… a lot of our patients with cardiotoxicity also have metabolic syndrome and these blood sugar issues.} 
\end{quote}}
\added{The ability to track these metrics in real-time continuously allows for the early detection of signs such as arrhythmias or blood pressure fluctuations, which may not be apparent during routine clinical visits.}

\added{\textbf{Opportunity 2: Remote Monitoring for Geographically Distant Patients.}}
\added{Participants highlighted the advantage of using wearables in providing remote access to vital health data for patients who are geographically distant from the clinic or unable to visit regularly.}
\added{As P1 explained:
\begin{quote}
    \pquote{P1}{A lot of our patients... travel very long distances to come here for treatment... they may call and be having an event where they feel really lightheaded or like they almost passed out... Having that data, especially if they can’t come in right away, is really helpful.} 
\end{quote}}

\added{This opportunity responds to \textbf{Challenge 2} by enabling remote, real-time monitoring through wearables, allowing clinicians to track patient health during gaps in clinical interactions and addressing symptoms that might otherwise go unnoticed.}

\added{Extending this opportunity, clinicians highlighted the need for patient-specific baselines that can alert the care team only when significant deviation occurs, rather than bombarding them with excessive data. This need informs our \textbf{Design Revision 2: Patient-Specific Baseline Alerts for Remote Monitoring.} For example, a sudden increase in the
heart rate that deviates from a patient’s established pattern could trigger an alert for further investigation, as P6
suggested, \textit{``the care team really should only be alerted if there's a significant change. That would be (when) something is wrong, we would need to intervene. Otherwise, it's just too much data not useful.''}}

\added{\textbf{Opportunity 3: LLM-Based Conversational Agent for Symptom Tracking}}
In addition to the potential of wearables in capturing physiological data, participants also saw great potential in using LLM-based CAs to track patient-reported symptoms over time without requiring manual input from patients. 
Several participants mentioned that this could ease the burden on patients, many of whom struggle to remember or document their symptoms accurately between appointments.
P2 emphasized the convenience of voice interaction, especially for patients who may have difficulty with traditional reporting methods, such as filling out forms, ``\textit{This is a really good idea. And that way they talk, and they don't need to be typing or spending time.}'' 
Furthermore, clinicians pointed out that combining wearables and LLM-based CAs could be particularly useful, as each technology addresses different aspects of patient monitoring. 
These two modalities complement each other and offer the ability to link physiological vital signs with patient-reported symptoms, as P1 commented, \textit{``being able to correlate the time with their biometrics could be really helpful.''}
\added{This opportunity potentially addresses \textbf{Challenge 3} by providing automated symptom tracking, reducing reliance on patient memory, and improving the accuracy of symptom reporting through the correlation of physiological and self-reported data.}

\added{Building on this opportunity, clinicians first noted the importance of monitoring critical symptoms such as chest discomfort, palpitations, and shortness of breath, which are typical early indicators of cardiotoxicity. As this need is closely related to \textbf{Design Revision 1}, we now extend it to \textbf{Monitor Key Metrics and Symptoms}. Secondly, as clinicians emphasized the importance of correlating symptoms with physiological data, they expressed a clear need for dynamic and flexible visualization of patient information, combining high-level summaries with detailed data views. This leads to our \textbf{Design Revision 3: Enhanced Visualizations for Critical Data Interpretation.} This dual approach would enable
quick assessments while also allowing deeper dives into specific trends and variations when necessary. For instance, a summary graph of daily average heart rates can offer a snapshot of a patient’s cardiovascular status. At the same time, the option
to view hourly data would provide insight into specific fluctuations and their potential causes. P5 emphasized the importance of trends on isolated data points, stating, ``\textit{The trend is extremely more important than just the one time.''}
This flexibility in visualizing data enables clinicians to identify patterns and correlations that might not be immediately
obvious from a single metric. For example, being able to drill down into hourly heart rate data can help clinicians
understand the context of a daily average that deviates from the baseline, facilitating more informed decision-making.}

\added{\textbf{Opportunity 4: AI-based Summarization and Predictive Risk Scores}}
Several participants noted that navigating extensive medical records could be time-consuming and challenging, \added{echoing \textbf{Challenge 4}}, and pointed out that AI-powered summarization tools can potentially help alleviate this burden by providing concise overviews of patient records. P4 echoed this sentiment by highlighting the growing issue of clinician burnout, much of which is driven by the need for documentation in EHRs: \textit{``There's a lot of burnout in medicine, and the burnout is from all of this documentation that we do. I see a huge role of AI in that documentation... fulfill out the EHR data automatically.''}

Also, participants saw promise in AI-generated predictive tools, particularly for assessing cardiotoxicity risk. 
As mentioned in Section ~\ref{sec3.2}, clinicians currently rely on a limited set of risk factors to determine which patients may be at higher risk. 
Participants emphasized the lack of standardized tools for predicting cardiotoxicity and expressed their need for such tools: ``\textit{I have not come across any risk scoring for cardiotoxicity. So I think that would be helpful.}''
Participants also noted that AI-driven risk scores could be dynamic and adjust as patient data are updated, offering them a continuous risk assessment. 
For example, P7 described how AI could assist in this ongoing process:
\begin{quote}
 \pquote{P7}{I think with AI there will be tools in place to even suggest what the next order should be... AI would help you understand what the next step could be even before you start talking to the patient.}
\end{quote}

In addition to its potential, clinicians stressed the importance of the explainability of AI models. They expressed the need for AI models to clearly communicate what risk scores mean and what actions should follow. Without explanations, they may be unsure how to act on the AI's predictions. 
For instance, P1 discussed the kind of explanation needed: \textit{``like the trigger event of having this higher score is maybe consideration for cardiology referral.''} The need for explainability is also closely tied to how clinicians perceive AI in their workflows. 
For many participants, they expressed the same sentiment that AI tools should serve as assistants rather than replace their clinical judgments.

\added{This need informs our \textbf{Design Revision 4: Clear Explanation of Technology to Prevent Misunderstanding}. 
While we presented AI modules and LLM-based voice assistants to clinicians, clinicians expressed their expectations that they should be more explainable to prevent misunderstandings and ensure proper usage. 
For example, understanding the basis of AI risk scores is crucial for clinicians to trust and effectively use the system. 
As P4 commented, 
\begin{quote}
    \pquote{P4}{So you have to see what you build it (AI risk score) on... How will you define the score? The problem with scores is that you have to sit down and manually do that. If something does it for us, we'll be happy to use it. ...AI is supposed to be my assistant, not my replacement.}
 \end{quote}}
\added{They would like to know how these scores are calculated and what they present.} 
\added{Also, they mentioned the need for LLM-based voice assistants to provide clear explanations about how AI responds to different patient scenarios. 
As P8 pointed out, understanding when to escalate care based on AI alerts is crucial: \textit{``Shortness of breath, discomfort, and passing out should always prompt contact with the healthcare provider... something that is acute, that is changing, should be evaluated by a healthcare professional.''}}

\section{System Prototype}
\label{sec:4}
\begin{figure*}[!h]
    \centering
    \includegraphics[width=1\textwidth]{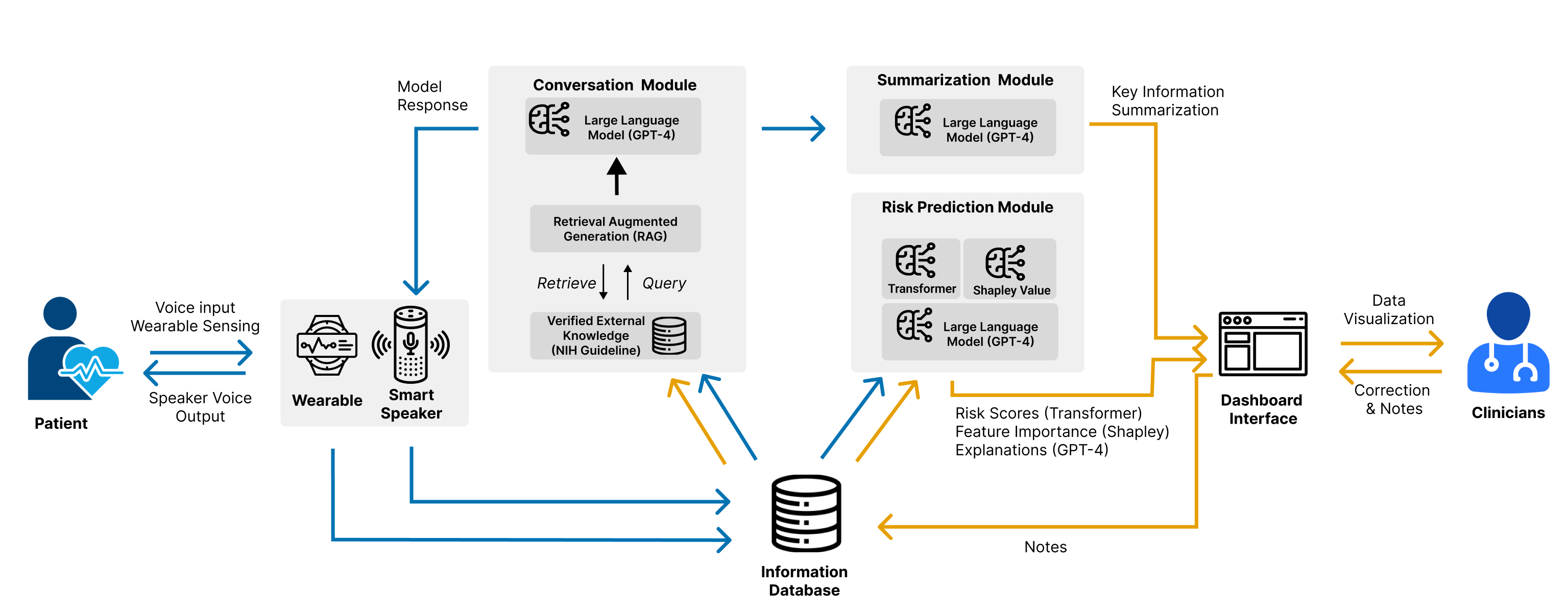}
    \caption{System Architecture of CardioAI. It integrates a wearable and a smart speaker to continuously collect physiological data and patient-reported symptoms. Data is processed by three key LLM-powered backend components: a conversation module, a summarization module, and a risk prediction module. The processed data, including key health summaries and cardiotoxicity risk scores with explainability, is then visualized on a clinician-facing dashboard.}
    \Description{This figure illustrates the system architecture for a cardiotoxicity monitoring solution that integrates multimodal data, such as wearable sensor data and voice input, with AI-powered modules for summarization, risk prediction, and conversation. On the left side of the diagram, the patient interacts with the system by providing voice input through a wearable device and smart speaker. The voice input and wearable data are sent to the system, which consists of three primary modules. The first module is the Conversation Module, which uses a large language model (GPT-4) and a Retrieval Augmented Generation (RAG) component that retrieves verified external knowledge (such as NIH guidelines). The retrieved data is used to generate responses for the patient. The second module is the Summarization Module, also powered by GPT-4, which summarizes key information from the data. The third module is the Risk Prediction Module, which uses a Transformer model and Shapley values to predict cardiotoxicity risk and explain the important factors influencing the prediction. All data is stored in an Information Database, and a Dashboard Interface is available to care providers, displaying the summarized data, cardiotoxicity risk scores, feature importance (as determined by Shapley values), and allowing for the inclusion of notes and corrections. The entire system is designed to provide continuous monitoring and decision support for both patients and clinicians, leveraging AI models to assess risk and aid in early diagnosis.}
    \label{fig:system}
\end{figure*}

\added{The opportunities and suggested design revisions outlined in Section ~\ref{sec3.2.2} guided the development of our pilot CardioAI system, designed to address the four key challenges in managing cancer treatment-induced cardiotoxicity. The system employs a multi-modal AI-based approach, offering continuous remote monitoring and explainable risk prediction to support clinicians' decision-making. 
An overview of our pilot system architecture is shown in Figure ~\ref{fig:system}. The pilot system is designed to seamlessly collect patient data from wearable devices and an LLM-VA~(Section ~\ref{sec4.1}), process it through the backend infrastructure~(Section ~\ref{sec4.2}), and present collected information and explainable AI risk scores on a clinician-facing dashboard~(Section ~\ref{sec:4.3}).} 

\subsection{Wearable and Smart Speaker Hardware}
\label{sec4.1}
Our findings in Section ~\ref{sec3.2.2} suggest opportunities~\added{(\textbf{Opportunity 1, 2 \& 3})} to use wearables and LLM-VA to monitor patient physiological data and patient-reported symptoms continuously.

The \textbf{wearable device} of our system is the Garmin Vivosmart 5~\cite{garmin:vivosmart5}, selected for its capability to continuously collect and monitor physiological signals suggested by clinicians, including heart rate, respiration rate, blood oxygen saturation, and skin temperature. 
As clinicians have expressed the importance of minimal effort for patients to use the device, we designed this subsystem to ensure data collection remains unobtrusive. 
Physiological data are collected every 10 seconds and periodically transmitted to the study app on the patient's smartphone.
The study app installed on the patient's device runs passively in the background, coordinates the data sync with the Garmin wearable without any user interaction, and securely uploads the collected data to the system's backend. The inclusion of the \textbf{smart speaker} in our system, supported by LLM, allows patients to self-report symptoms through natural spoken interactions. We chose Amazon Echo Dot\footnote{\url{https://www.amazon.ca/Echo-Dot-5th-Gen/dp/B09B8V1LZ3}} as our smart speaker device as it is affordable and easy to use. It comes equipped with built-in Speech-to-Text and Text-to-Speech functionality, and Alexa Skills Kit integration\footnote{\url{https://www.amazon.ca/b?ie=UTF8&node=16286269011}} ensures a smooth connection to the backend. 

\subsection{Backend Algorithms}
\label{sec4.2}
The backend architecture stores, processes, and analyzes the data the wearable and smart speaker collects with the following modules. 

\paragraph{Conversation Module}
The Conversation Module processes patient input from the smart speaker and enables dynamic, personalized interactions through natural language processing. The core of this module is powered by a Retrieval-Augmented Generation (RAG)~\cite{lewis2020retrieval, gao2023retrieval} LLM, based on OpenAI's GPT-4 model~\cite{openai_platform, openai2024gpt4technicalreport}, which is hosted on Microsoft Azure’s HIPAA-compliant infrastructure~\cite{openai2024gpt4technicalreport}. 
We chose the RAG approach for its ability to enhance LLM’s generative capabilities by retrieving domain-specific knowledge from external resources, such as research articles on cancer treatment-induced cardiotoxicity and established clinical guidelines~\cite{lyon20222022}.
Furthermore, the backend is designed to retrieve and reference prior conversations with the patient. For instance, if a patient previously reported palpitations, the system can refer to that symptom in future conversations, asking whether the palpitations have changed in frequency or severity since the last report. 

\begin{figure}
    \centering
    \includegraphics[width=0.5\textwidth]{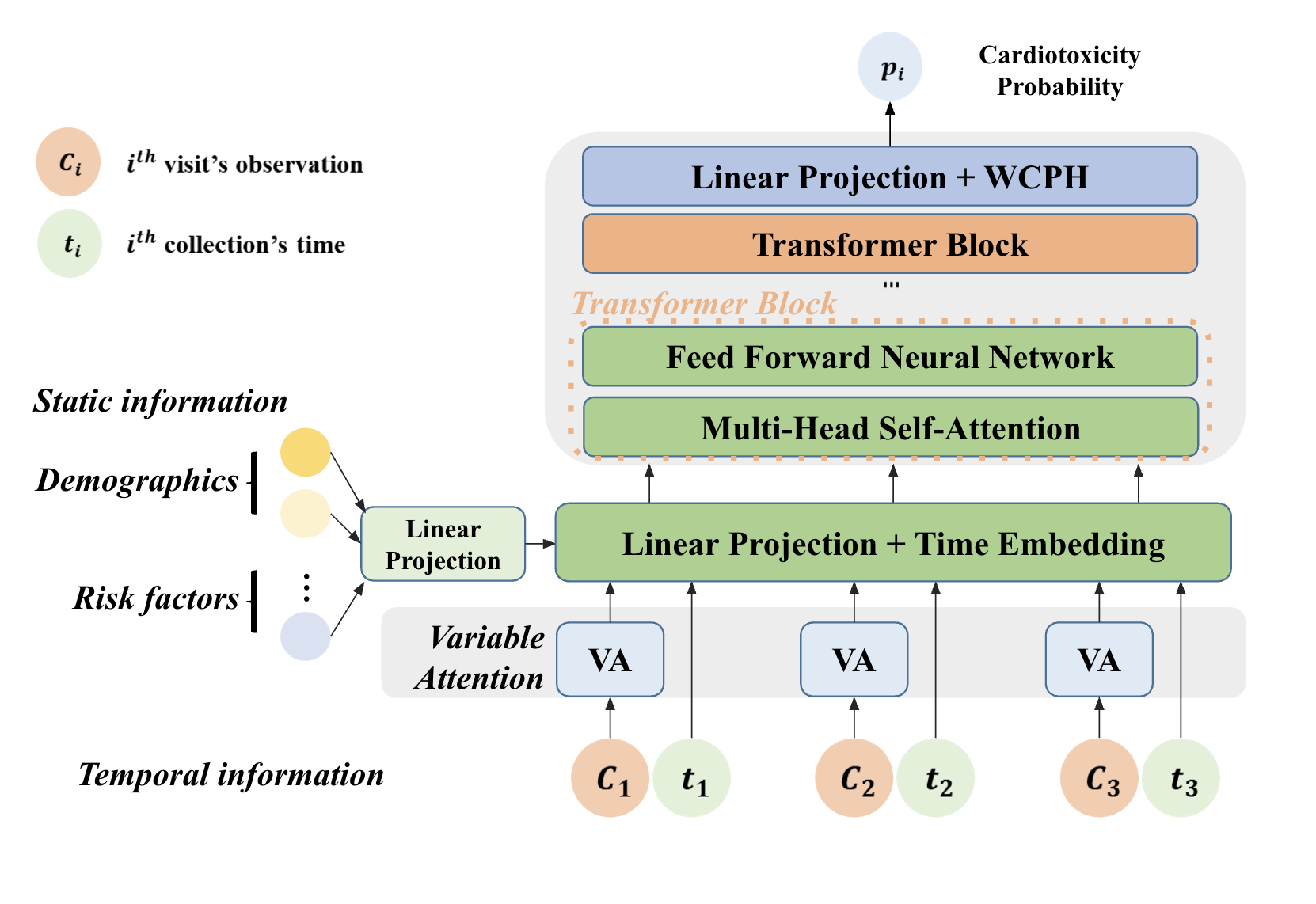}
    \caption{AI-based Cardiotoxicity Risk Score Prediction Framework.}
    \Description{The figure illustrates the architecture of the cardiotoxicity prediction framework. The framework processes static and temporal patient data to predict cardiotoxicity probability. Static information, such as demographics and risk factors, undergoes linear projection. Temporal information, including observations and collection times, is processed through a variable attention (VA) mechanism and a time embedding layer. The outputs from these layers feed into a transformer block, which comprises multi-head self-attention and a feed-forward neural network. The final output of the transformer block is combined with a linear projection layer and a Weibull Cox proportional hazards (WCPH) model to generate the cardiotoxicity probability.}
    \label{fig:Transformer}
\end{figure}

\paragraph{Risk Prediction Module}
\added{We leverage Transformer~\cite{vaswani2017attention} as the backbone to predict cardiotoxicity after assignment to treatment. 
A logistic regression model is used first to analyze long-term EHRs to identify high-risk factors with positive coefficients. These identified risk factors, alongside 
static patient information (e.g., demographics), and temporal data (e.g., prior medical events), are passed into the Transformer model to calculate the probability of cardiotoxicity risk, as \autoref{fig:Transformer} shows. 
The Transformer model processes sequences of patient visits and generates health state vectors, embedding medical codes, procedures, and medications into fixed-size low-dimensional vectors to mitigate high-dimensionality issues common in EHR data. 
To focus on the most relevant medical events within a visit, we integrate a variable attention module, which automatically prioritizes critical events for accurate cardiotoxicity prediction.
For time-to-event prediction, we deploy a Weibull Cox proportional hazards (WCPH) model, converting the cardiotoxicity risk prediction into a survival analysis framework. 
The risk score is calculated as a product of a baseline hazard function and the patient-specific cardiotoxicity risk based on the Transformer’s output, reflecting the clinical trajectory of patients over time. 
All learnable parameters, including the scale and shape parameters of the Weibull distribution~\cite{barrett2014weibull}, the hazard weights, and Transformer parameters, are optimized through a mini-batch stochastic gradient descent process~\cite{eon1998online}. 
To improve the interpretability of the model, we adopt the Shapley value method~\cite{winter2002shapley}, providing clinicians with clear information on the key factors driving the predictions of high or low cardiotoxicity risk. 
Additionally, we incorporate an LLM-based explanation generation module to provide plain-language explanations of risk scores and Shapley values, making the results more accessible to clinicians.}

\paragraph{Information Database}
We employ a cloud-based database to store collected data from the wearable and smart speaker, conversational logs, and historical EHRs. These data will facilitate the functionality of the other modules in the system. The database is encrypted to protect each individual’s privacy.

\paragraph{Summarization Module}
The Summarization Module generates comprehensive daily summaries by retrieving and synthesizing data from Information Database: patient EHRs, previous conversation logs, and physiological data. This module, powered by GPT-4~\cite{openai_platform}, ensures that healthcare providers receive a clear, concise, and holistic view of the patient’s health status each day.
By pulling data from the Information Database, the module reviews historical EHRs to provide essential context, the most recent conversational data, and data from the wearables with a particular focus on detecting any abnormalities or deviations in physiological metrics such as heart rate, respiration, or SpO2 levels.

\subsection{Frontend: Information Dashboard UI}
\label{sec:4.3}
\added{Based on the opportunities and design revisions suggested by clinicians in Section~\ref{sec3.2.2}, we transformed the initial low-fidelity UI prototype into a detailed and functional final interactive dashboard. While the initial iteration explored multiple design and content options, the final UI reflects an agreed-upon choice that consolidates clinicians' preferences into a streamlined, user-centered interface, which operationalizes the conceptual representations into practical and cohesive elements tailored to meet clinicians’ workflows and decision-making needs.}

\added{Specifically, we reorganized the overall layout to position wearable sensor data and conversation logs on the same page, placing them in parallel, which allows clinicians to view physiological metrics alongside patient-reported symptoms simultaneously without additional navigation. In addition, we prioritized displaying the metrics and symptoms that clinicians identified as most relevant for cardiotoxicity management and adopted visualization formats that clinicians suggested as most appropriate for each metric. To Design Revision 4, we revised the section at the upper-right corner of the dashboard, which defines the cardiotoxicity risk score, includes dynamic visualizations of risk scores over time, and incorporates feature importance indicators to clarify the key factors contributing to the score. Furthermore, we separate the summary of symptoms and metrics from the prior summary module and locate this section below the patient information, which leverages LLM-generated summaries to highlight abnormalities. 
}
The final dashboard interface has five main modules, as shown in Figure~\ref{fig:teaser}. 

\textbf{Module A: Patient Information Overview} integrates seamlessly with existing EHR systems, providing immediate access to critical patient information without the need to switch between different platforms, such as demographics, cancer type, stage, and treatment history.

\textbf{Module B: Daily Summary} aggregate and summarize key information from collected self-reported symptoms and wearable sensor data in a concise format to support a quick overview of the patient's current status, particularly focusing on any significant changes or issues that might require immediate intervention. 

\textbf{Module C: Wearable Sensor Data} displays the collected physiological data, such as heart rate, respiration, oxygen saturation, and skin temperature. The visualization is presented in an interactive format, allowing providers to explore different data modalities over specific time periods, quickly identifying trends or anomalies in the patient's physiological signals. 

\textbf{Module D: Explainable AI-generated Risk Score} presents the predicted cardiotoxicity risk score and its trends. We also present its Shapley Value as a breakdown of the most influential factors that predicted the risk score and an LLM-generated plain-language explanation of the risk score and its Shapley Value. 

\textbf{Module E: Conversation Log} shows the raw conversation log between the patients and the smart speaker. The visualization leverages a clear color scheme to differentiate ``normal'' (green) and ``abnormal'' (yellow) symptoms reporting and a selection feature to enable clinicians to select specific dates to review the detailed conversation logs.

\section{Evaluation}

\added{To understand clinicians’ perceptions of our pilot system, we conducted a heuristic evaluation study with clinicians to interact with and provide feedback on our pilot system. Each individual evaluation focused specifically on assessing the clinician-facing dashboard~(Section ~\ref{sec:4.3}) to systematically examine how clinicians would utilize the system for decision-making for cancer treatment-induced cardiotoxicity, further addressing RQ2. The evaluation results demonstrate that participants generally appreciated CardioAI’s simplicity, ease of use, improved access to relevant information, and support for proactive clinical decision-making.}

\subsection{Methods: Participants and Procedure}

\added{As we focus on evaluating how the pilot system would support clinicians' decision-making at this phase, our heuristic evaluation served as a design probe to explore specific aspects of the clinician-facing dashboard, such as the presentation of information, accessibility, clinicians’ information usage, and their overall perception. The goal was to refine the system’s design by evaluating its usability and identifying areas for improvement, leveraging targeted insights from a smaller and highly specific group of clinicians before engaging a larger cohort of clinical experts in future evaluations.
We adopted the same methodology described in Section ~\ref{sec3.1} to recruit clinicians from our previous participatory design study cohort. Through convenience sampling, we recruited four clinicians~(P3, P5, P7, P11): two oncologists and two cardiologists. This balanced participant group was selected to ensure high sample specificity~\cite{malterud2016sample}, representing the primary specialties involved in cancer treatment-induced cardiotoxicity management. Given the study’s narrow aim of evaluating the clinician-facing dashboard’s usability and decision-making support, this specific and well-defined group provided sufficient information power~\cite{malterud2016sample} to address our research objectives.
Before the study, we pre-loaded synthetic patient data into the prototype, representing cardiotoxicity scenarios typically encountered in clinical practice. 
At the beginning of each session, researchers first introduced the purpose of the study and provided a brief explanation of the think-aloud methodology. 
Clinicians then engaged in unstructured interaction with the dashboard. They freely navigated its features and explored its functionalities while verbalizing their thought processes and actions. To observe natural interaction patterns, researchers refrained from providing specific guidance during this phase.}

Following each think-aloud session, we then conducted a semi-structured interview with each participant. \added{The interview questions can be found in Appendix ~\ref{appendix:evalQ}. }
These interviews \added{further} explored how they envisioned using the \added{dashboard} for decision-making, their \added{perceptions} of the usability in a clinical setting, and their expectations or concerns about integrating the system into their workflow. The study concluded with a brief evaluation questionnaire including the NASA Task Load Index (TLX)~\cite{hart1988development} and the System Usability Scale (SUS)~\cite{Brooke_1995_SUS}, to assess perceived workload and system usability.
The TLX~\cite{hart1988development}questionnaire used a 7-point Likert scale to assess different elements of the task load. For questions related to demand, effort, or frustration, lower scores reflected a lighter workload, which was preferred for the system. For performance questions, higher scores indicated better results. The SUS~\cite{Brooke_1995_SUS}, a widely recognized usability scale, asked users to rate their agreement with 10 statements, ranging from "strongly disagree" to "strongly agree", on topics like ease of use of the system, the need for technical support, and learning difficulty.
Two researchers processed the interview transcriptions with thematic analysis, as in the previous participatory design sessions. We summarize our findings below.

\subsection{Findings}

\subsubsection{Clinicians' Uses of Information for Decision-making}
During think-aloud sessions, clinicians described how they would use the information in their decision-making, and we identified how the system could support them in the three stages of the current decision-making workflow as discussed in related work: symptom identification, diagnostic testing and collaboration with other specialties, and clinical decision-making and intervention.  

\paragraph{Symptom Identification} In the first stage of the diagnosis of cardiotoxicity, clinicians often begin by identifying symptoms based on patient self-reports, either outside the clinic or during in-person visits. 
Clinicians praised the system's streamlined access to symptom reports via the voice assistant module. They typically examined the conversation log for detailed information about when and how symptoms were reported. 
Clinicians highlighted the importance of correlating these reported symptoms with the patient’s physiological data, and they appreciated the system's ability to seamlessly align symptoms with physiological data for deeper insight. 
For example, when patient-reported symptoms such as chest pain or shortness of breath were present, clinicians would immediately check the wearable module for specific physiological metrics like heart rate, blood pressure, and blood sugar levels. 
As P7 described his thinking process: \textit{``What was the symptom? So chest pain, shortness of breath... the question is what was the heart rate, what was the blood pressure? What was the blood sugar?''}

Also, clinicians appreciated the system’s ability to provide quick and easy access to crucial metrics, such as blood pressure and heart rate, which they consistently check during patient assessments. P3 emphasized the importance of these vital signs in their routine evaluations, stating: ``\textit{Blood pressure and heart rate are the key things walking into the room, like any and every time.}''

\paragraph{Diagnostic Testing and Collaboration with Other Specialties}
After identifying potential symptoms of cardiotoxicity, clinicians typically move to the diagnostic stage, where they review recent test results or order new tests to confirm or rule out cardiotoxicity. Clinicians praised our system for streamlining the retrieval of diagnostic data, allowing them to easily access past and current test results in one place and interpret diagnostic results in conjunction with the patient’s medical history without switching between different platforms or systems.
Clinicians frequently utilized the "Results" and "Therapy" tabs to review key diagnostic tests for heart function, such as echocardiograms, cardiac MRIs, and EKGs, which are crucial for assessing cardiotoxicity. 
As P3 emphasized: \textit{``For the heart laboratory tests, the first things I look at are the cardiac laboratory tests... like did they have an echocardiogram, a cardiac MRI, or even EKGs.''} Beyond diagnostic results, clinicians valued how the system integrates a patient’s treatment history with their diagnostic data.

\paragraph{Clinical Decision-Making}
Once the diagnostic evaluation is complete, clinicians need to decide whether to adjust the patient’s treatment plan, initiate new interventions, or increase monitoring to mitigate cardiotoxicity risk. 
Clinicians emphasized the value of our system’s real-time cardiotoxicity risk scores to guide these critical decisions. The dynamic nature of the risk scores provided a clear picture of the risk of cardiotoxicity of a patient over time.

Clinicians appreciate that the system allows them to assess cardiotoxicity risk before proceeding with treatment. P3 highlighted how this feature supports informed decision-making, particularly when choosing between treatments with varying levels of risk: 
\begin{quote} \pquote{P3}{If you could at baseline, incorporate all the information that you have on the patient and say, ‘Hey, this patient's actually super high risk for cardiotoxicity from treatment option A, but treatment option B is less risky,’ then that could actually inform your treatment.} 
\end{quote} 
By offering this comparison, clinicians could balance the potential risks of cardiotoxicity against the efficacy of cancer treatments, enabling more tailored and safer treatment plans.

In addition to the initial risk assessment, clinicians valued the ability to track changes in the cardiotoxicity risk score over time. This feature enabled them to detect increasing risks and intervene before serious cardiac events occurred. 
This capacity for ongoing risk monitoring allowed clinicians to dynamically adjust treatment plans, ensuring that patients were protected from severe cardiotoxicity while still receiving effective cancer care.

Moreover, clinicians appreciated how easily accessible and interpretable the risk scores were, which contributed to making fast, informed decisions without being overwhelmed by excessive data. P11 expressed their satisfaction with the system’s straightforward design, ``\textit{This risk score is nice. I would look at that quickly... It doesn’t look cumbersome.}'' This simplicity allowed clinicians to integrate the risk score into their workflow seamlessly, supporting their decision-making process efficiently and effectively.

\subsubsection{Usability of Our System}

During the study, clinicians were able to interpret the presented information and incorporate it into their existing workflow. 
As they interacted with the system, clinicians highlighted several key aspects of the system's usability and its potential to support their clinical decision-making. The average SUS score is $72.33\pm1.89$ (out of 100), suggesting it offers a solid level of usability.  

\paragraph{Simplicity and Ease of Use}
A major theme highlighted by participants was the system's \textbf{simplicity and ease of use}, which they considered essential for smooth integration into their clinical workflow. All participants rated the system very highly for \textit{ease of use} in SUS, indicating that they found the system very easy to navigate, such as P3's comment: \textit{``like this interface so far, is relatively simple... It doesn’t look cumbersome.''}
This simplicity in design was further supported by the NASA-TLX effort score, which averaged $1.13\pm0.70$ (out of 7.0), indicating the system required minimal effort to operate and could be adopted easily. 

Participants also expressed high \textbf{confidence} in their ability to use the system effectively, supported by the SUS confidence metric, and many stated that they felt comfortable completing tasks without the need for additional help. They agreed that most people would be able to \textbf{learn to use the system very quickly}, a sentiment captured by the high SUS score for this question, highlighting its low learning curve and easy fit within their existing workflows.

\paragraph{Improved Access to Information and More Relevant Information}

Participants emphasized that the system's \textbf{ease of access to relevant information} was a significant advantage. By presenting the data in a streamlined manner, the system minimized the need to search through multiple records manually, improving workflow efficiency. As P11 explained, \textit{``If you have more information, this will make it easier without us having to look back in the chart.''}
Furthermore, the NASA-TLX results highlighted that the system did not impose a high workload on users as mental demand ($1.13\pm0.23$) and physical demand ($0.73\pm0.42$) are low. This suggests that the system effectively supported the retrieval and organization of information without imposing an additional burden on users’ mental or physical resources.

Participants also appreciated the system’s ability to deliver ~\textbf{highly specific and organized information}. This targeted presentation of data was considered essential to help them make informed decisions related to cardiotoxicity. P11 pointed out that the system effectively prioritized cardiotoxicity-related information with a more focused view: \begin{quote} \pquote{P11}{In the chart, there's so much more data. This is more specific, organized, mostly for cardiotoxicity. So the app provides that more specific concerns information.} \end{quote}

\paragraph{Supporting Proactive and Informed Clinical Decision-Making}

The system's capacity to ~\textbf{facilitate proactive decision-making} was a significant benefit highlighted by the participants. By providing real-time updates and predictive insights, the system allowed clinicians to anticipate potential complications and intervene early, shifting from a reactive to a more proactive approach. P5 remarked,
\begin{quote}
    \pquote{P5}{I think workflow wise, this would actually just preempt us ... so at least this way, we have a little bit more data to say like, hey, you should go.. or hey, just follow up with your cardiologist... I think that part would be really helpful in terms of how we actually manage patients and what we refer them.}
\end{quote} In addition to enhancing proactive care, the system was praised for its role in ~\textbf{complementing existing clinical tools} rather than replacing them. Clinicians felt that the system could augment their capabilities to monitor patients. 

Clinicians also recognized the AI-based system’s potential for ~\textbf{supporting long-term monitoring and follow-up care}, particularly for patients in the survivorship phase of cancer treatment. The system’s ability to continuously monitor patient health for extended periods was seen as a valuable tool to detect later-emerging cardiotoxic effects and ensure ongoing patient safety. P3 suggested, \begin{quote}
    \pquote{P3}{Maybe in your survivorship, you know your patients who are 5, 10 years out from their doxorubicin or their chest radiation... having a 30-day monitoring period before their surveillance visit would help make sure there’s no worrying signs there that would prompt more cardiac.}
\end{quote}

\subsubsection{Future Expectations}
During think-aloud sessions, clinicians also provided their future expectations for the system, highlighting where they envision the prototype could be further improved. 

\paragraph{Enhancing Data Relevance and System Integration}

While clinicians highlighted the system's improved access to relevant information as a significant advantage, they also expressed the need for more clinically relevant data to support decision-making further. One key request was for comprehensive tracking of all patient medications, including non-cancer treatments that could interact with cancer therapies or exacerbate cardiotoxicity. P7 pointed out, ``\textit{This is only the cancer treatment. Normally, it's not just 3 or 4 medications... there are other medications},'' highlighting the complexity of managing patients undergoing cancer treatment. By incorporating a more thorough medication history, the system could better support clinicians in recognizing potential drug interactions and side effects that could impact patient health outcomes.

Clinicians also expressed the need for more detailed information about patients' anti-cancer treatments, such as the type of chemotherapy administered, its last administration date, cumulative dose, and associated cardiotoxic risks. They emphasized that not all anti-cancer treatments have the same cardiotoxic potential, and understanding these specifics is essential for effective management. For example, P7 remarked, ``\textit{I would want to know what exactly is the chemotherapy that we're talking about, because not all chemotherapies have cardiotoxicity},'' pointing to the necessity of providing granular details to assess risks accurately and make informed treatment decisions.

Furthermore, while clinicians praised the system's ability to streamline access to relevant cardiotoxicity-related information, reducing the need to search through multiple records, they also expressed the expectation for it to integrate more seamlessly with the tools they already use. Reflecting on past experiences, they described the challenge of navigating between various systems and managing the multiple risk scores and alerts already in EHRs. As P5 noted, 
\begin{quote}
    \pquote{P5}{We do have, in our electronic medical record, a lot of these risk scores that are already kind of popping up and showing up for us. There’s a risk score for opiate abuse, a risk score for hospital readmission, a couple of different things. So we want these tools, but there’s also some alarm fatigue. Sometimes we see so much when we log in that we just kind of ignore it and get to what we need to do in the chart.}
\end{quote}

\paragraph{Customization Based on Clinicians' Specialties}
Another prominent theme from the feedback was the collaborative nature of cardiotoxicity decision-making, emphasizing the need for the system to offer customization based on the user’s specialty. 
Decision-making in cancer treatment-induced cardiotoxicity typically involves multiple specialists, such as cardiologists and oncologists, who each prioritize different aspects of patient data. For instance, cardiologists are primarily concerned with cardiac metrics, while oncologists need detailed information about anti-cancer treatments. P7, a cardiologist, highlighted this divergence in priorities, explained, ``\textit{For a cancer doctor, the mammogram is useful, important. But for me, it probably doesn't give any additional information,}'' pointing out the varying relevance of certain data types depending on the specialist's focus. In contrast, P5, a medical oncologist, stressed the need for a quick, high-level overview of cumulative anthracycline dosage, reflecting their focus on cancer treatment-related metrics.

In addition, the collaborative nature of decision-making often requires the input of both cardiologists and oncologists, making it crucial for the system to facilitate information sharing between specialties. P11 highlighted this collaborative dynamic, emphasizing the importance of reviewing notes from other specialists to gain a comprehensive understanding of the patient’s condition: 
\begin{quote}
    \pquote{P11}{I do look at the recent progress notes. I mean, sometimes the patient forgot what happened. So it's also good for me to go and look at those notes before walking in. So I know what other providers have seen. And then I can also estimate what issues I should address even if the patient doesn't bring it up, because the cancer team may have issues that they want some level of input on. I spent a lot of time looking these notes to try to understand what's going on. So we can make decisions about whether it's the cancer drugs or not.}
\end{quote}
By catering to the specific needs of different specialists while enabling collaboration, the system can make it easier for clinicians to access the information most relevant to their respective roles and share critical insights across specialties.

\subsubsection{Concerns}
\label{sec:concerns}
\paragraph{Ethical Concerns}
While remote monitoring systems have the potential to improve patient care through real-time data collection, clinicians expressed ethical concerns around their practical implementation, especially regarding timely interventions, ambiguity in responsibility, and the lack of patient education. 

One significant concern is the gap between real-time data availability and timely interventions. Although the system could alert clinicians to a patient's acute medical condition, there remains uncertainty about how quickly these alerts can be acted upon, especially outside regular office hours. Clinicians expressed worry about how the system would function when they are not readily available to respond. 
As P5 explained,
\begin{quote}
    \pquote{P5}{I'll be honest that it's gonna be hard to have, or to even remember to look at this dashboard sometimes, right? Or even if you check it 8 am every morning, you're gonna miss, you're inevitably gonna miss something later or you'll catch it later after the fact.}
\end{quote}

A further concern related to timely intervention is the ambiguity of responsibility when critical alerts are generated. Clinicians expressed that, even when alerts are issued promptly, the uncertainty around who is responsible for acting on them adds another layer of complexity. As P11 pointed out, ``\textit{identifying who's gonna take charge if there's a red flag that comes out is a significant challenge.}'' P7 expanded on this concern by posing critical questions about how these alerts would be managed in practice, given the constraints of staffing and availability: 
\begin{quote} \pquote{P7}{Thinking about staffing, thinking about like, who's going to be checking in with the patient? Obviously, if the patient just starts talking, there's no guarantee that someone's available to listen and take that message. Because if the healthcare system is being notified that the patient has a acute medical issue like, how does that get to the right person to figure out, hey, do they need to call 911? Or do we need to get them to the emergency room right now?} \end{quote} This highlights the logistic challenge of ensuring that the right person is available to receive the alert and empowered to make critical decisions based on it. 

Another recurring theme in the discussions was the risk of patients placing too much trust in the system, assuming that entering data into an app or chatbot would suffice for urgent medical issues. Clinicians were concerned that misplaced trust can lead patients to delay seeking immediate medical attention, relying solely on the chatbot for guidance. This points to a critical need for patient education as P11 emphasized the importance of clear communication with patients when they need to take action: 
\begin{quote} \pquote{P11}{We have to make sure that patients know... Some patients might think, 'Oh, well, I told my chatbot that I had chest pain, so it’s fine'... but they need to know what red flag symptoms are and that they need to seek immediate medical attention.} \end{quote} 
P7 further stressed that patients need to be aware of the system's limitations during off-hours, ``\textit{Patients need to know that if they told their chatbot that they had chest pain at 9 pm on a Friday… no one's gonna check the chatbot until Tuesday when they get back.}''

\paragraph{Accuracy and Validation of Information}
Clinicians also raised concerns about the accuracy and validation of the information. They were cautious about relying on data generated by technologies without rigorous validation. P7 noted: 
\begin{quote} \pquote{P7}{If it is not (validated), how should I be sure that this is all accurate information there? Because if I’m not convinced, like I might be looking at the information, but if I’m not sure whether the information is accurate, then we'll go back to the patient's chart in our own system and look for that information.} \end{quote} 
Clinicians expressed that without confidence in the accuracy of the data, they are inclined to verify it against existing systems, which potentially undermines their efficiency. P10 echoed this concern, emphasizing that for new technologies to gain adoption in clinical settings, they need to undergo the same rigorous validations as existing medical tools.

\section{Discussion}

In this discussion, we explore the broader implications of our findings. We begin by discussing the potential of our system in extending clinicians' capabilities beyond clinical settings (Section ~\ref{sec6.1}). We then propose design considerations for AI-assisted proactive decision-making (Section ~\ref{sec6.2}). We highlighted the technical, ethical, and privacy challenges that arise from implementing these AI-based systems, with particular attention to infrastructure, patient reliance, and data security concerns (Section ~\ref{sec6.3}). Finally, we presented the limitations and future directions of our work~(Section~\ref{sec6.4}). 

\subsection{Blurring the Boundary Between Clinical and Non-Clinical Settings} 
\label{sec6.1}
In this paper, we shed light on the critical challenges of decision-making for cancer treatment-induced cardiotoxicity. One of the overarching factors underlying these challenges is the traditional separation between clinical and non-clinical settings, which shapes how monitoring, diagnosis, and treatment decisions are made. 
Historically, healthcare delivery has been episodic and predominantly confined to structured environments such as clinics and hospitals, where clinicians rely on standardized tools and real-time data available only during patient visits~\cite{committee2001crossing}. This hospital-centric care model, however, is increasingly inadequate for conditions like cardiotoxicity, where symptoms can manifest unpredictably and outside scheduled clinical visits. 

The vision of extending healthcare beyond traditional clinical settings is not new. Over decades, various models and frameworks have been advocated for a more integrated, continuous approach in healthcare~\cite{campbell1998integrated, cueto2004origins, holter1961new, perednia1995telemedicine, jackson2013patient}. For instance, the Patient-Centered Medical Home (PCMH)~\cite{jackson2013patient, stange2010defining, rittenhouse2009patient} promotes comprehensive care coordination and management that spans various care environments. Similarly, the integration of telemedicine and RPM technologies ~\cite{catalyst2018telehealth, us2008national, kofoed2012benefits} has enabled continuous patient monitoring in non-clinical settings, helping to address chronic and high-risk conditions, such as heart failure, diabetes, and cancer treatments~\cite{darkins2008care}. Researchers in healthcare and the HCI community have been advancing this vision by designing technologies that allow clinicians to remotely track patient data~\cite{vegesna2017remote, 8857717, dias2018wearable, hardcastle2020fitbit, gresham2018wearable, tadas2023using} and facilitate patient communications~\cite{qiu2021NurseAMIE, gregory2023exploring, zhou2024performance, li2024scoping, yang2024talk2care, xu2021chatbot, kim2024mindfuldiary}. 

Our work builds on these studies by specifically operationalizing these visions in the context of cancer treatment-induced cardiotoxicity. While previous research has demonstrated the promise of RPM and CDSS, the unpredictable nature of cardiotoxicity necessitates a more proactive and continuous approach. Our findings suggest that \system, a multimodal AI-based system, has the potential to extend clinicians' capabilities beyond clinical settings by providing continuous monitoring of symptoms and real-time risk prediction. By integrating wearable devices, LLM-based voice assistants, and AI-driven analysis, our system facilitates seamless information collection from non-clinical settings into clinical workflows. Our study offers a practical application that 'blurs the boundaries' between clinical and non-clinical settings. 

While our study demonstrates the promise of multimodal AI-based systems in extending clinical capabilities, we also underscore the importance of considering the broader implications of these systems. Continuous data streams can be invaluable for conditions like cardiotoxicity, where early intervention is critical, but they may pose challenges in other contexts, such as causing patient anxiety or information overload in less dynamic or chronic conditions. These technologies should not be applied uniformly; instead, we argue that their use needs to be tailored to specific healthcare contexts. Rather than attempting to erase the boundary between clinical and non-clinical settings, we advocate for thoughtfully extending clinical capabilities to better address the needs of clinicians and patients, particularly for conditions that benefit from early detection and intervention. This perspective acknowledges the strengths of traditional clinical settings while recognizing the need for flexibility in delivering care across diverse environments.

\subsection{Design Considerations for AI-Assisted Proactive Decision-Making}
\label{sec6.2}

\added{Our studies highlight the potential of CardioAI in supporting clinical decision-making for cancer treatment-induced cardiotoxicity. Clinicians emphasized that integrating continuous monitoring with predictive risk scores facilitates a proactive approach to managing cardiotoxicity, enabling earlier interventions and more informed decisions before critical conditions arise. This represents a significant shift from traditional reactive workflows, aligning with broader trends in healthcare toward preventive care, as discussed in prior work~\cite{wagner1998chronic, darkins2008care}, and demonstrates how our system translates this approach into actionable clinical practice.}

\added{Additionally, we propose several design considerations critical for the successful deployment of our system in real-world settings. These insights also serve as valuable guidance for future designers developing multimodal AI-based systems for clinical applications.} 

\added{\subsubsection{Fostering Accurate User Expectations through UI and XAI Design.} A foundational design consideration for multimodal AI-based systems in clinical settings is fostering accurate user expectations by leveraging UI and explainable AI (XAI) design. Given the complexity of these systems and their deployment in high-stakes clinical environments, it is essential to communicate their roles, capabilities, and limitations to users. We suggest that such communication be seamlessly embedded within the system's design to ensure accessibility and usability while mitigating risks of over-reliance or inappropriate use.}

\added{A key aspect of fostering accurate user expectations is to provide explicit delineation of the system's roles, capabilities, and limitations. Misunderstandings about what the system can and cannot do can lead to misuse with potentially severe consequences. For example, clinicians in our study emphasized the importance of ensuring that patients understand the LLM-VA’s intended role as a tool for symptom tracking and not as a substitute for emergency care. Misuse in such situations could lead to delays in critical interventions, which may pose significant risks to patient safety. We suggest that designers use UI and XAI design strategies to integrate this information directly into the system, instead of relying on methods like training sessions, manuals, and tutorials, which may not always be practical or accessible. For instance, designers can embed tooltips within the interface to provide contextualized definitions and explanations of the system functionality and limitations and use visual cues, such as color-coded indicators, to highlight critical outputs. For example, in our system, AI-based predictive risk scores were accompanied by detailed definitions, and red color was used to signify the detection of red-flag symptoms to alert clinicians about high-risk scenarios.}

\added{In addition to defining system roles and limitations, we recommend that designers provide interpretable and actionable explanations of the system's outputs. Clinicians in our study highlighted the need to understand how AI-based risk predictions are generated, what they signify, and how they can inform actionable insights. In response to this need, our system incorporates explainable risk scores and feature importance indicators, which outline the key factors influencing each prediction. Additionally, these risk scores are accompanied by actionable recommendations tailored to different risk levels, providing clinicians with clear guidance on potential next steps. We suggest that designers consider how to provide explanations tailored to the nature of the technologies used in the system, such as detailing the sources of outputs, clarifying the reasoning behind predictions, and presenting insights in a concise and actionable format.}

\added{Explainability and expectation management should also remain dynamic, evolving alongside technological advancements and clinical workflow changes. For example, when new features are added or algorithms are improved, interfaces should be updated to reflect these changes accordingly. This could better support ongoing dialogue between stakeholders and designers, fostering a collaborative relationship that builds trust and accountability over time.}

\added{\subsubsection{Balancing Clinical Benefits and Added Workload.}
Another design consideration for multimodal AI-based systems in clinical decision-making is ensuring that the systems deliver clinical benefits without imposing additional workload and cognitive burden on clinicians. While the potential of such systems lies in their ability to provide comprehensive information, their effectiveness hinges on how this information is curated, prioritized, and presented. Poorly designed systems could overwhelm users with excessive information, reduce efficiency, and hinder clinicians' ability to make timely and informed decisions. }

\added{To achieve additional clinical benefits, such systems often require the collection of extra data or the introduction of new information into clinicians' workflows. For example, in our study, clinicians highlighted the tension between the value of having more data and the potential for information overload. Integrating AI-generated risk scores into the system provided valuable insights but also required clinicians to spend additional time interpreting these scores, assessing their accuracy, and evaluating their relevance. This inherent trade-off between added benefits and increased cognitive effort presents a significant design challenge. To address this, we recommend that designers leverage implicit user interactions with newly introduced information or functionalities as feedback to improve the system, taking this as an opportunity to minimize the additional workload introduced. This approach aligns with AI-in-the-loop principles, enabling systems to adapt dynamically to user behaviors. For instance, clinicians' frequent engagement with specific flagged metrics could guide the system to refine its prioritization algorithms, thereby reducing cognitive effort and enhancing efficiency over time.}

\added{Another key aspect is designing systems that align with the diverse workflows and priorities of different clinical specialties. For instance, cardiologists in our study prioritized EKG data as a critical diagnostic tool, while oncologists placed greater emphasis on treatment plans and patient histories. This divergence underscores the importance of customizable system features, such as tailored notifications, adaptable dashboards, and role-specific thresholds. We recommend that designers actively collaborate with clinicians during the design process to identify these specialty-specific needs and ensure that systems accommodate them effectively. However, customization must be balanced with consistency: while personalized features can enhance relevance, preserving core functionalities across all user groups is essential to maintaining system usability, reliability, and safety.}

\subsection{Technical, Ethical and Privacy Concerns}
\label{sec6.3}

\added{While the scope of our study focuses on designing an interactive UI for the clinician-facing information dashboard to support clinicians' decision-making, which does not extend to patient-facing interfaces or technologies, clinicians highlighted several concerns regarding the multimodal AI-based system for clinical decision-making. These concerns encompass infrastructure challenges, ethical considerations, and privacy risks which need to be addressed before real-world deployment.}

\subsubsection{Infrastructure Challenges}
 A critical concern is the need for robust infrastructure support, including reliable internet connections, functioning devices, and interoperability between systems. Clinicians noted that technical failures, such as malfunctioning devices or poor Wi-Fi connectivity, could disrupt real-time symptom monitoring, delay interventions, and undermine the effectiveness of the system. Ensuring reliable IT infrastructure and technical support will be critical for the success of these systems. In addition to infrastructure, human resources are essential for ongoing monitoring and maintenance. While AI can automate data collection and analysis, human oversight remains crucial for interpreting alerts and responding to abnormal data. Clinicians expressed concerns about the increased workload that this can place on healthcare teams, particularly in 24/7 monitoring environments. Managing these demands will require careful planning to avoid overburdening staff and ensure continuous, high-quality care.

\subsubsection{Ethical Concerns}
Beyond technical and resource challenges, clinicians raised ethical concerns, particularly around the potential for patients to over-rely on AI-based systems. Clinicians expressed concern that patients might assume logging symptoms into a system guarantees immediate intervention, even when healthcare teams are unavailable, without fully understanding the limitations of technology. This risk is further amplified in underserved communities where access to clinicians is already limited, potentially leading to a false sense of security.

In this context, the role of patient education becomes paramount. Our findings suggest that empowering patients through technology must be accompanied by clear communication about the system's capabilities and limitations. While AI-based systems can provide valuable insight and suggestions, patients need to be aware of which symptoms warrant immediate medical attention and which can be monitored through the system. Without this education, there is a risk that patients misunderstand the urgency of their symptoms and rely on the system inappropriately, leading to delayed interventions. This underscores the importance of designing AI-based systems that not only provide accurate information but also guide patients in making informed decisions about when to escalate their concerns to human clinicians.

\subsubsection{Privacy and Security Cocerns}
Privacy and security concerns are also central to the implementation of AI and LLM-based systems. As these technologies handle sensitive patient data, ensuring the confidentiality and integrity of this information becomes crucial. Our study revealed that clinicians are particularly concerned about how patient data is stored, transmitted, and accessed within AI-based systems. The increasing reliance on AI raises questions about data breaches and unauthorized access, which could undermine patient trust and pose significant legal risks. Clinicians also highlighted the need for transparency in data handling practices to build trust among stakeholders. This includes clear communication about who has access to the data, how it is used, and the safeguards in place to protect it. Robust data governance frameworks are essential to address these concerns, ensuring compliance with legal and ethical standards while maintaining the confidentiality and integrity of patient information. While this study focuses on designing the clinician-facing dashboard for decision support, addressing privacy and security concerns across the entire multi-modal AI-based system -- including patient-facing components such as wearables and voice assistants -- will be critical. These aspects fall outside the scope of the current work but will be addressed in future efforts to design and integrate a complete remote monitoring and decision-making support system. 

\subsection{Future Directions \& Limitations}
\label{sec6.4}
\added{
Our work has several limitations. First, our sample sizes are limited. We involved 11 clinicians during the participatory design phase and four clinicians for the heuristic evaluation, similar to prior studies using small, expert-focused samples to gather domain-specific insights~\cite{yang2024talk2care,sepsis,cai2019hello,jacobs2021designing, beede2020human}. 
Recruiting clinicians, especially in specialized domains like cardiotoxicity, presents unique challenges due to their demanding schedules, limited availability, and the niche nature of the expertise required. Despite these constraints, we considered the concept of information power, which posits that smaller, targeted samples can be sufficient when participants are highly specific to the study’s aims and the data collected is rich and relevant~\cite{malterud2016sample}.
The recurring themes and consistent patterns observed across both phases further support the adequacy of our sample sizes in achieving the study’s objectives. 
However, all participants were recruited from a single hospital, which may introduce systematic biases. Also, our participant pool only included cardiologists and oncologists -- the primary stakeholders in cardiotoxicity decision-making -- other specialists, such as rheumatologists, radiologists, and nurses, could offer valuable perspectives depending on the specific cancer types or treatment pathways. We suggest future work could expand the participant pool to include a more diverse range of roles and institutions to enhance the generalizability of findings further and explore the system’s applicability in different clinical contexts.}

\added{Second, our system is a proof-of-concept prototype and has not been integrated into EHR systems or deployed in real-world clinical settings. The current prototype was used as a design probe to elicit clinician feedback on usability and decision-supporting functionalities, serving as a preliminary feasibility evaluation in a simulated environment. By conducting this initial evaluation, we aimed to avoid prematurely investing significant time and effort from domain experts in a large-scale controlled usability experiment before addressing key design and functional concerns. While this approach provides valuable early insights, it may influence the validity and generalizability of our findings. As discussed in Section~\ref{sec6.3}, real-world deployment may reveal additional challenges, such as seamless EHR integration, operational scalability, and privacy and ethical concerns. Future work should consider these challenges to ensure the system’s robustness and practical utility.}

\added{Third, our study specifically focused on the interactive UI design of the clinician-facing dashboard and its core functionalities from the clinicians’ perspectives. We explored how the clinician-facing dashboard could support clinicians’ decision-making in managing cancer treatment-induced cardiotoxicity. Evaluations of patient-facing components, such as patients’ interactions with the LLM-VA or their experiences using wearables for symptom tracking, were beyond the scope of this study. Future work should include comprehensive and systematic evaluations of these patient-facing modules, such as assessing the LLM-VA for symptom reporting and wearables for continuous monitoring. These aspects are critical for scaling the system to real-world deployments and ensuring it effectively meets the needs of all stakeholders, including both patients and clinicians. }

\added{Also, fostering trust in AI-based tools is fundamental, particularly in high-stakes clinical settings. In this study, we explored the potential of explainable AI technologies to help domain experts better understand the outputs of AI and LLM-based systems. However, trust is a multifaceted concept, encompassing not only technical transparency but also reliability, openness, tangibility, etc. Future work could delve deeper into strategies for fostering trust in LLM-VA and AI models, such as incorporating uncertainty metrics that highlight the confidence level of predictions, and explore the interplay between these strategies and users' perceptions of trust.}

\added{Last but not least, while this study focused on cardiotoxicity, there is a broader need to investigate how similar AI-based systems can be applied to other clinical areas where early detection and intervention are critical. By exploring the deployment of proactive AI tools in various healthcare contexts, future work can help improve patient outcomes and optimize decision-making processes across different conditions.}

\section{Conclusion}

In this study, we explore clinicians' challenges in managing cancer treatment-induced cardiotoxicity and develop a multimodal AI-based system, CardioAI, to support symptom monitoring and risk prediction. Through participatory design sessions with 11 clinicians, we uncovered the complexity of managing cardiotoxicity and identified gaps in existing monitoring approaches. Our system provides continuous monitoring of symptoms via wearable devices and LLM-based VA, and explainable AI-based predictive risk scores, which offer actionable insights to support clinical decision-making. Evaluation by clinical experts highlighted the system’s ability to reduce information overload, streamline workflows, and support proactive decision-making. These findings contribute to the growing field of HCI in healthcare, providing a foundation for future development of technologies to address similar challenges in clinical decision-making tasks. We envision that our work can inspire future designs of multimodal AI-based systems. 

\begin{acks}

This work was funded in part by the National Science Foundation under award number IIS-2145625, and IIS-2302730, by the National Institutes of Health under award number R01AI188576 and R01MD018424, and by The Ohio State University President’s Research Excellence Accelerator Grant and by The Northeastern University Tier-1 Research Grant. The content is solely the responsibility of the authors and does not necessarily represent the official views of the National Science Foundation or the National Institutes of Health.
The authors extend heartfelt thanks to the participants from The Ohio State University Wexner Medical Center.

\end{acks}

\bibliographystyle{ACM-Reference-Format}
\bibliography{ref}

\appendix

\added{
\section{Participatory Design Exit Interview Script}
\label{appendix:interviewQ}
\begin{itemize}
    \item \textbf{Question 1 - Background: }Could you please tell me a bit more about yourself and your practice, such as your years of study, residence, daily workload, etc.?
    \item \textbf{Question 2 - Experience of Cancer Treatment-Induced Cardiotoxicity: }Can you recall a recent cardiotoxicity encounter for a cancer patient during treatment or near the completion stage? When did it happen? What types of cancer and cancer treatment were given? 
    \item \textbf{Question 3 - Diagnosis and Monitoring of Cancer Treatment-Induced Cardiotoxicity: }When do you suspect cardiotoxicity? What symptoms or signals lead to your suspicion? How do you solve the problem? 
    \item \textbf{Question 4 - Technology Uses and Potentials in Current Cancer Treatment-Induced Cardiotoxicity Management: } What kind of technology have you been using? Are you aware of any technologies or tools that others in your field or institutions have been exploring or adopting for cancer treatments and cardiotoxicity management? 
    \item \textbf{Question 5 - Perspectives Towards Potential Technologies: }Are you familiar with or have you explored the potential uses of emerging technologies, such as wearables, chatbots, or AI-based risk prediction models? Do you think any of these technologies could benefit risk management and early diagnosis? Do you have any concerns? [What information do you wish AI could have provided you?][What metrics or symptoms do you think might be important to collect when a patient is at home?]
    \item \textbf{Question 6 - Closing Questions:} Is there anything else that you would like to share with us or any questions you have for us?
\end{itemize}
}

\added{
\section{Heuristic Evaluation Exit Interview Script}
\label{appendix:evalQ}
\begin{itemize}
    \item \textbf{Question 1 - Overall Feedback: }Can you share your overall feedback on the system?
    \item \textbf{Question 2 - Information Use: } Do you find the information provided helpful for assisting in decision-making? Was the amount of information appropriate (e.g., too much, too little, or just right)?
    \item \textbf{Question 3 - Interaction Design: } How would you describe your overall experience with the system's user interface and interaction design? What about data visualizations? [further ask in terms of intuitive design, information display, customization, navigation, and alerts]
    \item \textbf{Question 4 - Concerns: } Do you have any concerns about using this system in clinical workflows?
    \item \textbf{Question 5 - Future Design: } What improvements or additional features would you suggest for future iterations of the system? [e.g., Is there any additional information you wish had been included?]
\end{itemize}
}

\onecolumn

\begin{table}[h!]
\centering

\caption{Qualitative Codebook of Participatory Design Study Findings (Part 1)}
\label{Appendix:codebook}
\resizebox{\textwidth}{!}{%
\begin{tabular}{p{0.19\textwidth}|p{0.19\textwidth}|p{0.6\textwidth}}
\toprule
\textbf{Theme} & \textbf{Sub-Theme} & \textbf{Example} \\
\midrule

\textbf{Challenges in Cardiotoxicity Decision-making}
& \multirow{3}{=}{Symptoms Not Just Subtle and Infrequent, But Sometimes Absent in the Early Stages}  
& "I got this message he had a severe form of arrhythmia... He was completely asymptomatic. He was sitting at home relaxing, and the monitor picked up this thing." (P4) \\
\cmidrule{3-3}
& & "So it all depends. If the patient, symptomatic or not, if they're asymptomatic, then we're the ones picking it up." (P6) \\
\cmidrule{3-3}
& & "If a toxicity arises, especially that they're asymptomatic from, they're very hesitant to want to reach out.." (P1) \\
\cmidrule{2-3}

& \multirow{2}{=}{Logistical Barriers Compound Self-Reporting Limitations}
& "I think as much as we encourage them to reach out to us if they're having symptoms, many of them still feel like they're bothering us if they call." (P2) \\
\cmidrule{3-3}
& & "In some areas, patients may not hear back right away from their physicians or nurses if they’re having symptoms... They may feel like they’re complaining too much or being a burden." (P1) \\
\cmidrule{2-3}

& \multirow{2}{=}  {Non-Specific Tools Compound Self-Reporting Limitations}
& "It was it kind of presented itself (cardiotoxicity-related symptoms), and it had been almost 4 or 5 days before when she had first described the symptom." (P2) \\
\cmidrule{3-3}
& & "We use a survey tool to ask patients how they’re feeling, but it’s not specific to cardiotoxicity.." (P5) \\
\cmidrule{2-3}

& \multirow{2}{=} {Absence of Risk Patterns as an Additional Burden to Clinician Workload}
& "What type of features correlate with the likelihood somebody's going to have cardiotoxicity is still, I think, very shaky. From what I understand, I think we don't have a good grasp." (P1) \\
\cmidrule{3-3}
& & "Because he had no cardiac risk factors. He had no heart risk factors that could predict this side effect." (P4) \\
\midrule

\textbf{Opportunities}
& \multirow{2}{=} {Continuous Monitoring of Key Clinical Metrics Using Wearables} 
& "I think we can pick up more patients with this. Or we can pick up patients earlier with this... We may be able to find more cardiac patients or cardiac adverse events related to our medications by doing this kind of monitoring." (P4) \\
\cmidrule{3-3}
& & "I don't know what they are like outside of the clinic ... Those things can sometimes be helpful. It's almost like people who have cardiac monitors." (P1) \\
\cmidrule{2-3}

& \multirow{2}{=}{Remote Monitoring for Geographically Distant Patients}
& "A lot of our patients... travel very long distances to come here for treatment... they may call and be having an event where they feel really lightheaded or like they almost passed out... Having that data, especially if they can’t come in right away, is really helpful." (P1) \\
\cmidrule{3-3}
& & "Doing more for patients in rural areas could be a place where this could work pretty well." (P11) \\
\cmidrule{2-3}

& \multirow{2}{=}{LLM-Based Conversational Agent for Symptom Tracking}
& "I asked them to keep checking at home and write it down on paper, and they bring it to me... But sometimes they forget, or you know things happen... the concept is great, and having the patient have control over the symptoms and speaking and just translated that into something that we can see." (P3) \\
\cmidrule{3-3}
& & "This is a really good idea. And that way they talk, and they don't need to be typing or spending time." (P2) \\
\bottomrule
\end{tabular}
}
\end{table}

\begin{table}[h!]
\centering
\caption{Qualitative Codebook of Participatory Design Study Findings (Part 2)}
\label{tab:cardiotoxicity_challenges_2}

\begin{tabular}{p{0.19\textwidth}|p{0.19\textwidth}|p{0.6\textwidth}}
\toprule
\textbf{Theme} & \textbf{Sub-Theme} & \textbf{Example} \\
\midrule

\textbf{Opportunities}
& \multirow{1}{=}{AI-based Summarization}
& "There's a lot of burnout in medicine, and the burnout is from all of this documentation that we do. I see a huge role of AI in that documentation... fulfill out the EHR data automatically." (P1) \\
\cmidrule{3-3}
& & "If you could summarize the medical record, I think that'd be really helpful. Instead of going all over the place, looking around, it would be helpful." (P2) \\
\cmidrule{2-3}
& \multirow{1}{=}{AI-based Risk Prediction} 
& "If you are able to do it upfront, it would be very helpful... you will have to talk to some cardiologists to see how you define the score." (P4) \\
\cmidrule{3-3}
& & "I think with AI there will be tools in place to even suggest what the next order should be... AI would help you understand what the next step could be even before you start talking to the patient." (P7) \\
\cmidrule{3-3}
& & "I have not come across any risk scoring for cardiotoxicity. So I think that would be helpful." (P5) \\
\midrule

\textbf{Design Suggestions}
& \multirow{2}{=}{Monitoring Key Clinical Metrics} 
& "Oxygen levels would be nice. I know a lot of these wearable devices can do that." (P2) \\
\cmidrule{3-3}
& & "It's heart rate. It's basically heart rate. Yeah, that can be monitored." (P4) \\
\cmidrule{2-3}

& \multirow{2}{=}{Monitoring Key Clinical Symptoms} 
& "Chest discomfort is good, shortness of breath, palpitations... These are all symptoms that we are looking out for... If they’re symptomatic, they should be checked quickly." (P4) \\
\cmidrule{3-3}
& & "This is good; shortness of breath, palpitation, especially palpitations." (P8) \\
\cmidrule{2-3}

& \multirow{2}{=}{Patient-Specific Baseline Alerts for Remote Monitoring}
& "But there are normal ranges for all of these things, but I think the change is based on what the patient's baseline is. There's something called clinically meaningful change. That really depends on where the patient baseline is." (P6) \\
\cmidrule{3-3}
& & "I think it's most helpful to have a baseline because people come in with all variety of kind of where they were at before." (P1) \\
\cmidrule{2-3}

& \multirow{3}{=}{Enhanced Visualizations for Critical Data Interpretation} 
& "The trend is extremely more important than just the one time." (P5) \\
\cmidrule{3-3}
& & "Whenever the patient comes for their infusions, whenever they have the encounters with their physician. That could also be a good time to review these trends and alerts and see how they would change treatment decisions." (P8) \\
\cmidrule{3-3}
& & "I think that would be helpful if patients are having an event where they felt like they were gonna pass out, or they were having palpitations to them, being able to correlate the time with whatever their biometrics were at the same time can be helpful." (P1) \\
\cmidrule{2-3}

& \multirow{2}{=}{Clear Explanation of Technology to Prevent Misunderstanding} 
& "So you have to see what you build it (AI risk score) on... How will you define the score? The problem with scores is that you have to sit down and manually do that. If something does it for us, we'll be happy to use it. ... AI is supposed to be my assistant, not my replacement." (P4) \\
\cmidrule{3-3}
& & "Whenever we get this data and these scores, what do we do with it? ..especially patients are going to see cause. At the end of the day, it's just having somewhat clear guidelines on what score means." (P1) \\
\bottomrule

\end{tabular}
\end{table}

\end{document}